\documentclass{article}
\pdfoutput=1

\usepackage[utf8]{inputenc}

\usepackage{amsmath,fullpage,graphicx,color,amsfonts,url,multirow,amssymb,tabularx,setspace}
\usepackage{natbib}

\newcommand\independent{\protect\mathpalette{\protect\independenT}{\perp}}
\def\independenT#1#2{\mathrel{\rlap{$#1#2$}\mkern2mu{#1#2}}}

\newcolumntype{L}[1]{>{\raggedright\arraybackslash}p{#1}}
\newcolumntype{C}[1]{>{\centering\arraybackslash}p{#1}}
\newcolumntype{R}[1]{>{\raggedleft\arraybackslash}p{#1}}

\title{Causal inference and machine learning approaches for evaluation of the health impacts of large-scale air quality regulations}
\author{Rachel C. Nethery$^1$, Fabrizia Mealli$^2$, Jason D. Sacks$^3$, Francesca Dominici$^1$\\
\small{$^1$Department of Biostatistics, Harvard T.H. Chan School of Public Health, Boston, MA USA}\\
\small{$^2$Department of Statistics, Computer Science, Applications, University of Florence, Florence, Italy}\\
\small{$^3$National Center for Environmental Assessment, Office of Research and Development,}\\ \small{U.S. Environmental Protection Agency, Research Triangle Park, NC, USA}}
\date{}

\begin{document}

\maketitle

\begin{abstract}
We develop a causal inference approach to estimate the number of adverse health events prevented by large-scale air quality regulations via changes in exposure to multiple pollutants. This approach is motivated by regulations that impact pollution levels in all areas within their purview. We introduce a causal estimand called the Total Events Avoided (TEA) by the regulation, defined as the difference in the expected number of health events under the no-regulation pollution exposures and the observed number of health events under the with-regulation pollution exposures. We propose a matching method and a machine learning method that leverage high-resolution, population-level pollution and health data to estimate the TEA. Our approach improves upon traditional methods for regulation health impact analyses by clarifying the causal identifying assumptions, utilizing population-level data, minimizing parametric assumptions, and considering the impacts of multiple pollutants simultaneously. To reduce model-dependence, the TEA estimate captures health impacts only for units in the data whose anticipated no-regulation features are within the support of the observed with-regulation data, thereby providing a conservative but data-driven assessment to complement traditional parametric approaches. We apply these methods to investigate the health impacts of the 1990 Clean Air Act Amendments in the US Medicare population.

\noindent%
\end{abstract}

\section{Introduction}\label{s:intro}

In its 2011 cost-benefit analysis of the 1990 Clean Air Act Amendments (CAAA), the United States Environmental Protection Agency (EPA) estimates that the total direct costs of compliance in the year 2000 were nearly \$20 billion, and it anticipates that these costs will increase to over \$65 billion by 2020 \citep{epa2011benefits}. The CAAA is an expansion of the 1970 Clean Air Act, and it has prompted the enactment of numerous new regulatory programs, both at national and local levels, to ensure that pollution emissions limits are observed and that air quality standards are being met (see \cite{henneman2019air} for a summary). Hereafter, for simplicity we use the term CAAA to refer to the set of regulations put in place to adhere to the law. While the EPA estimates that the economic benefits of the CAAA dwarf the costs (estimated 2000 benefits \$770 billion, estimated 2020 benefits \$2 trillion), the increasing compliance costs call for continued evaluations of the effects of the CAAA using the most advanced, rigorous methods. Of particular interest to the public are the health impacts of the CAAA.

Traditionally, assessments of the health impacts of large-scale regulations, including assessments of the CAAA, have relied on the combination of simulated air quality models and pollutant-health exposure-response functions (ERF). These assessments have relied on various tools including the World Health Organization's AirQ+ software \citep{who2019airq} and EPA's Environmental Benefits Mapping and Analysis Program - Community Edition (BenMAP-CE) software \citep{epa2011benefits,sacks2018environmental}. For example, within BenMAP-CE, atmospheric chemistry modeling is first used to estimate concentrations of a pollutant on a grid across the area of interest in a post-regulation year under (1) the factual/observed scenario of regulation implementation and (2) the counterfactual/unobserved scenario of no regulation (hereafter referred to as factual and counterfactual exposures). The difference in the factual and counterfactual pollutant levels in each grid cell are used as inputs in the ERFs. The ERFs employ these differences along with the size of the exposed population and a health effect estimate, i.e. a linear model coefficient from a previously published epidemiologic study capturing the relationship between pollutant exposure and a health outcome, to estimate the number of health events prevented by the regulation-attributable changes in pollutant exposures in the specified year. This process is performed separately for each relevant pollutant-health outcome combination. For more detail on the traditional approach and the ERFs, see Section 1 of the Supplementary Materials.

Causal inference principles have historically been featured heavily in analyses of the health effects of short-term air pollution interventions, which can often be formulated as natural experiments. Larger, gradually-implemented regulatory actions are more complex because 1) the resultant changes in levels of multiple pollutants vary in space and time, 2) long term health trends may coincide with changes in air quality, and 3) time-varying confounding is likely \citep{van2008evaluating}. Only recently have causal inference approaches begun to be developed to address these issues. \cite{zigler2012estimating} and \cite{zigler2018impact} investigate the effect of a component of the CAAA, the National Ambient Air Quality Standard (NAAQS) non-attainment designations, on health in the Medicare population using a principal stratification approach. To our knowledge, factual and counterfactual exposures from air quality modelling software and causal inference methodology have not yet been integrated for regulation evaluation, despite their natural connection.

In this paper, we seek to estimate the number of health events prevented by the CAAA in a specific year using a unique approach that combines factual and counterfactual pollution exposures with observed health outcome data from the Medicare population (instead of relying on ERFs derived from previously conducted epidemiologic studies). In particular, we intend to answer the question ``How many mortality events, cardiovascular hospitalizations, and dementia-related hospitalizations were prevented in the Medicare population in the year 2000 due to CAAA-attributable changes in particulate matter (PM$_{2.5}$) and ozone (O$_3$) exposures in the same year?'' Because pollution exposures may continue to impact health beyond the year of exposure, we also wish to determine how many of each of these health events were prevented in the year 2001 due to the CAAA-attributable changes in pollutant exposures in the year 2000. To investigate this question, we utilize estimates of PM$_{2.5}$ and O$_3$ exposure levels across the US for the year 2000 under the factual with-CAAA scenario and under the counterfactual no-CAAA scenario. We combine them with zipcode level counts of mortality, cardiovascular hospitalizations, and dementia-related hospitalizations from Medicare in the years 2000 and 2001. Then we utilize the number of health events observed under factual PM$_{2.5}$ and O$_3$ levels to inform estimation of the number of health events that would have occurred under the counterfactual pollutant levels (the counterfactual outcome).

We introduce a causal inference framework that can be applied to evaluate the CAAA or any other large-scale air quality regulation. Reliance on counterfactual pollution predictions, which come from the EPA's cost-benefit analysis of the CAAA \citep{epa2011benefits}, enables us to conduct a causal inference investigation in this otherwise intractable setting where we never observe any data under the no-regulation scenario. The first novel feature of our work is the introduction of a causal estimand, which we call the Total Events Avoided (TEA) by the regulation. It is defined as the sum across all units of the difference in the expected number of health events under the counterfactual pollution exposures and the observed number of health events under the factual pollution exposures. We also lay out the corresponding identifiability conditions. The second novel aspect of this paper is in the development of 1) a matching approach and 2) a machine learning method for estimation of the TEA. Relying on minimal modeling assumptions, these methods use confounder-adjusted relationships between observed pollution exposures and health outcomes to inform estimation of the counterfactual outcomes. While we are seeking to estimate the same quantity estimated in traditional health impact analyses of regulations (the number of health events prevented in a year due to regulation-attributable changes in pollutant exposures that year), the statistical methods used are quite different. Our approach improves on the traditional one by: 1) defining the causal parameter of interest; 2) spelling out the assumptions needed to identify it from data; 3) relying on population-level health outcome data for estimation; 4) minimizing parametric assumptions; and 5) accounting for the simultaneous effect of multiple pollutants and thereby capturing any synergistic effects. However, to avoid extrapolation and heavy reliance on parametric modeling assumptions, our methods exclude some areas from the analysis, thereby producing conservative yet data-driven estimates of the health impacts of the regulation.

Matching is one of the most commonly used approaches to estimate causal effects \citep{ho2007matching, stuart2010matching}. Machine learning procedures have emerged more recently as a tool for causal inference \citep{hill_bayesian_2011,hahn2017bayesian,louizos2017causal}. Both have primarily been used to estimate average treatment effects in settings with a binary treatment, with recent limited extensions to the continuous exposure setting \citep{kreif2015evaluation,wu2018matching}. To our knowledge, neither approach has been used to do estimation in the context of a multivariate continuous treatment (the pollution exposures in our setting). To estimate the TEA, we develop both a matching method and an adaptation of the Bayesian Additive Regression Trees (BART) algorithm \citep{chipman2010bart} to accommodate multivariate continuous treatments.

In Section~\ref{s:data}, we discuss the air pollution and Medicare data that motivate our methodological developments. In Section~\ref{s:methods}, we formally introduce the TEA and identifying assumptions, and we present our matching and machine learning methods for TEA estimation. Section~\ref{s:sims} describes simulations conducted to evaluate the performance of these methods. In Section~\ref{s:app}, we apply these methods to investigate the health impacts of the CAAA. Finally, we conclude with a summary and discussion of our findings in Section~\ref{s:discuss}.

\section{Data}\label{s:data}
In this paper, we focus on the number of health events prevented due to CAAA-attributable changes in two major pollutants-- PM$_{2.5}$ and O$_3$. These two pollutants are known to have large health impacts. We have obtained state-of-the-art factual (with-CAAA) and counterfactual (no-CAAA) gridded PM$_{2.5}$ and O$_3$ exposure estimates for the continental US in the year 2000. PM$_{2.5}$ exposures (both factual and counterfactual) are measured in $\mu g /m^3$ and represent annual averages, while O$_3$ exposures are measured in parts per billion (ppb) and represent warm season averages. Our factual pollution exposure estimates come from so-called hybrid models which combine ground monitor data, satellite data, and chemical transport modeling to estimate exposures on a fine grid across the US. The factual PM$_{2.5}$ exposure estimates employed here are introduced in \cite{van2019regional} and are produced at approximately 1 km$^2$ grid resolution. Our factual O$_3$ exposure estimates were developed by \cite{di2017hybrid}, also at 1 km$^2$ grid resolution.

We employ the year-2000 counterfactual gridded PM$_{2.5}$ and O$_3$ exposure estimates for the continental US from the EPA's Second Section 812 Prospective Analysis (hereafter called the Section 812 Analysis), its most recent cost-benefit analysis of the CAAA \citep{epa2011benefits}. To produce these, gridded hourly emissions inventories were first created for the counterfactual (no-CAAA) scenario in 2000, representing estimated emissions with scope and stringency equivalent to 1990 levels but adjusted to economic and population changes in 2000. Note that this approach to creating counterfactual emissions inventories assumes that 1) in the absence of the CAAA, no new emissions regulations would have been implemented in the US between 1990 and 2000 and 2) in the absence of the CAAA, the scope and stringency of US emissions would not have increased. We anticipate that in reality these assumptions would slightly bias the emissions estimates in opposite directions; thus, we expect that they balance each other out to create a realistic counterfactual emissions scenario. The emissions inventories were fed into atmospheric chemistry modeling software to produce estimates of counterfactual annual average PM$_{2.5}$ at 36 km$^2$ grid resolution, and counterfactual warm season O$_3$ at 12 km$^2$ grid resolution. For more detail on the creation of the counterfactual exposure estimates, see \cite{epa2011benefits}. We note that factual exposure estimates were also produced for the EPA's Section 812 Analysis in an analogous manner to the counterfactual exposures. For a discussion of why we have chosen to use factual exposures from the hybrid models instead, see Section 2 of the Supplementary Materials.

\begin{figure}[h]
\centering
\includegraphics[scale=.31]{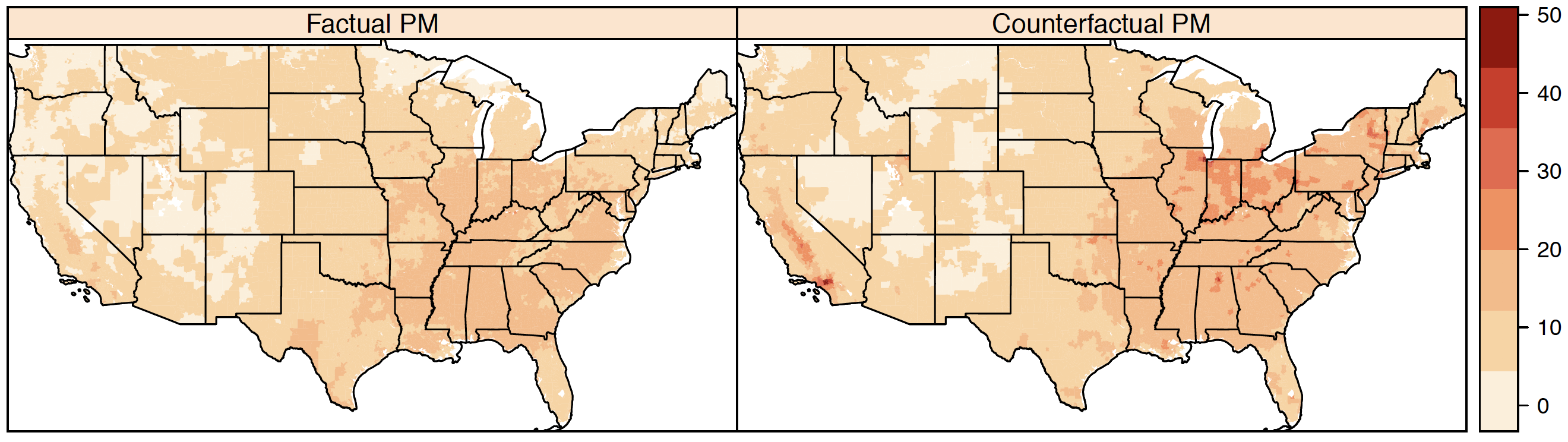}
\includegraphics[scale=.31]{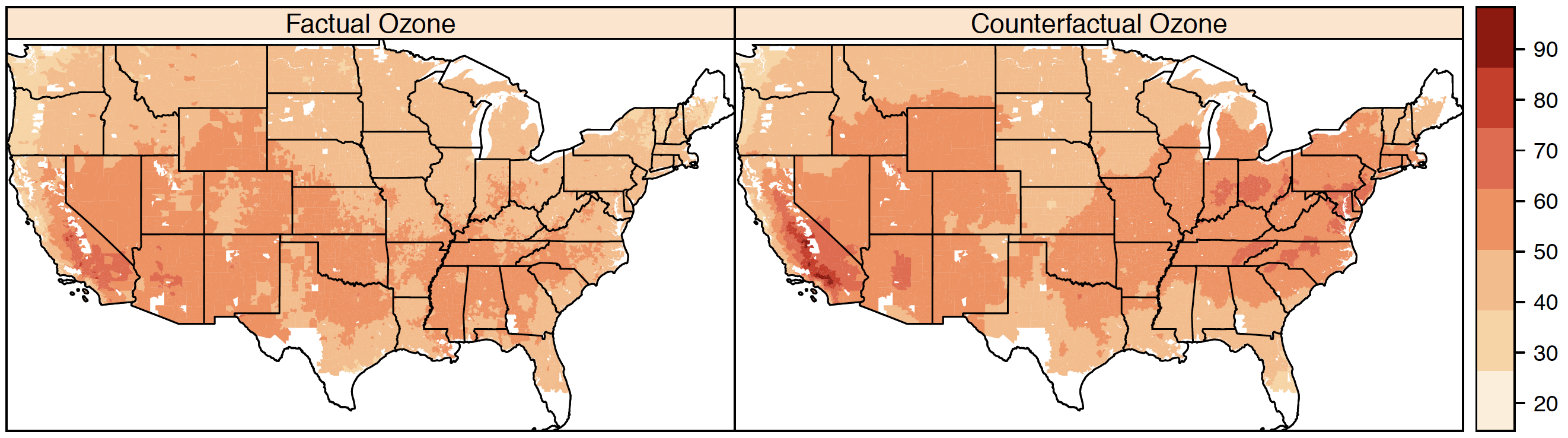}
\caption{Maps of estimated year-2000 zipcode level factual and counterfactual annual average PM$_{2.5}$ levels in $\mu g /m^3$ and warm season average O$_3$ levels in parts per billion (ppb). The factual pollutant values are estimates of the true exposures with the CAAA. The counterfactual values reflect the anticipated pollutant exposures under the emissions scenario expected without the CAAA.}
\label{fig:maps}
\end{figure}

We use area-weighting to aggregate the gridded PM$_{2.5}$ and O$_3$ values to zip codes, in order to merge it with the Medicare data. See the zipcode maps of the factual and counterfactual PM$_{2.5}$ and O$_3$ data in Figure~\ref{fig:maps}. Throughout the paper, we make the strong assumption that the observed and counterfactual pollution levels are known or estimated without error, because current limitations in air quality modeling impede reliable quantification of uncertainties \citep{nethery2019estimating}. This is a commonly used assumption in air pollution epidemiology, and we discuss the implications and potential impacts in Section~\ref{s:discuss}.

We construct datasets containing records for all Medicare beneficiaries in the continental US for the years 2000 and 2001. Medicare covers more than 96\% of Americans age 65 and older \citep{di2017air}. Because cross-zipcode moving rates are low in the Medicare population \citep{di2017air,awad2019change}, we make the assumption that, for each of the 2000 and 2001 cohorts, subjects were exposed to the year-2000 pollution levels of their zipcode of residence. Zipcode counts of mortality, cardiovascular hospitalizations, and dementia-related hospitalizations in each cohort will serve as the outcomes in our analysis. For detailed information about how these counts were constructed, including the ICD-9 codes used to classify hospitalizations, see Section 2 of the Supplementary Materials. We selected these health outcomes on the basis of previous literature associating each of them with air pollution \citep{pope2002lung,moolgavkar2003air,brook2010particulate,
power2016exposure}. 
We also compute the size of the Medicare population in each zipcode in each year so that rates of each health event can be computed (e.g. count of mortalities in the Medicare population in the zipcode in year 2000/total Medicare population in the zipcode in year 2000).

Finally, a set of potential confounders of pollution and health relationships is constructed. Confounder data come from the 2000 US census. All confounders are zipcode-aggregate features that reflect all residents of the zipcodes, not only Medicare beneficiaries. They are percent of the population below the poverty line (poverty), population density per square mile (popdensity), median value of owner-occupied properties (housevalue), percent of the population black (black), median household income (income), percent of housing units occupied by their owner (ownhome), percent of the population hispanic (hispanic), and percent of the population with less than a high school education (education).

\section{Methods}\label{s:methods}

\subsection{Estimand and Identifying Assumptions}\label{ss:estimand}
In this section, we propose a causal inference estimand to measure the difference in the number of health events in a given year under the regulation and no regulation scenarios, and we lay out the identifying assumptions. The units of analysis for our case study of the CAAA will be zipcodes, but throughout this section we use the term units to emphasize the generality of the approach to any areal units. Let $Y_i$ denote the count of health events observed in unit $i$, $i=1,...,N$, and $P_i$ the at-risk population size in unit $i$. Then let $Y^*_i=\frac{Y_i}{P_i}$ be the event rate. While our estimand involves counts, we must conduct modeling with rates instead of counts when at-risk population sizes vary. Let $\boldsymbol{X}_i$ denote a vector of observed confounders of the relationship between pollution and the health outcomes under study, as well as any factors besides the observed pollutants which were impacted by the regulation. Let $\boldsymbol{T}_i$ denote the $Q$-length vector of continuous pollutant exposure measurements for unit $i$, where $Q$ is the number of pollutants under consideration (annual/warm season averages), and we assume that $\boldsymbol{T}$ has compact support over a Q-dimensional hyperrectangle, $\mathbb{Z}$, which is a subspace of $\mathbb{R}^Q$. $\boldsymbol{T}_i$, the vector of pollutant exposure levels, will serve as the ``treatment'' variable in our causal inference framework.

We develop our approach within the potential outcomes framework of \cite{rubin1974estimating}. Recall that we will make use of both the factual and counterfactual pollution levels for each unit. Then let $Y_i(\boldsymbol{T}=\boldsymbol{t}_{1i})$ be the number of health events that unit $i$ would experience under its observed pollutant values ($\boldsymbol{t}_{1i}$) in the factual scenario of regulation implementation. Let $Y_i(\boldsymbol{T}=\boldsymbol{t}_{2i})$ be the number of health events that unit $i$ would experience under its counterfactual pollution levels ($\boldsymbol{t}_{2i}$) in the no-regulation scenario, with all other features of the unit identical to the observed ones. For each unit, $Y_i(\boldsymbol{T}=\boldsymbol{t}_{1i})$ is observed and $Y_i(\boldsymbol{T}=\boldsymbol{t}_{2i})$ is unobserved.  These potential outcomes can be constructed under the stable unit treatment value assumption (SUTVA) \citep{rubin1980randomization}. SUTVA requires that the health effect of a given treatment (pollution) level is the same no matter how that treatment level is arrived at, and that one unit's pollution level does not affect the number of health events in another unit (the later is known as the no interference assumption). The former assumption could be violated if, for instance, certain sources of pollution are more toxic than others. No interference is likely to be a strong assumption \citep{zigler2012estimating,zigler2018bipartite}, as most individuals are regularly exposed to the pollution levels in areas other than their area of residence, which could impact their health. More discussion of this assumption is provided in Sections~\ref{s:app} and~\ref{s:discuss}.

Before formalizing the TEA, we emphasize that the TEA is designed to quantify the health impacts of the regulation only through resultant changes in the pollutants in $\boldsymbol{T}$, while holding all other measured features fixed at the observed levels (under the regulation). The TEA characterizes a counterfactual scenario in which the only changes from the observed with-regulation scenario are the pollution exposures, and compares the number of health events in this counterfactual scenario and the observed one. We explain the motive behind this estimand. Previous causal inference analyses \citep{zigler2012estimating,zigler2018impact} have targeted the effects of air pollution regulations on health both via the resultant changes in pollutant exposures (associative effects) and via other intermediates (dissociative effects). These studies investigated an element of the CAAA, the NAAQS non-attainment designations, which impacted some locations and not others, providing variation in attainment status and observed health outcome data under both the attainment and non-attainment scenarios. These data enabled the estimation of both associative and dissociative effects. Here we instead wish to evaluate a larger regulation which had ``universal'' impacts in areas under its purview (hereafter referred to as a universal regulation). We do not observe any health outcome data whatsoever under the no-regulation scenario. With this limitation, our analysis must rely entirely on the observed outcome data under regulation. Because the only information we have about the no-regulation scenario is the counterfactual pollutant exposures, it is only through these pollutants that the observed health outcome data can inform estimation of the health outcomes in the no-regulation scenario. Thus these data do not allow any investigation of dissociative effects, which is why we have defined the TEA as the effect of the regulation only through the resultant changes in the pollutants in $\boldsymbol{T}$. The Section 812 Analysis also estimates health effects exclusively through regulation-attributable changes in pollutants. Previous studies have found positive dissociative health impacts of regulations \citep{zigler2012estimating}. If such effects exist, the TEA will under-represent the total number of health events prevented by the regulation.

Now invoking the notation laid out above, the number of health events prevented due to regulation-induced changes in pollutants is
$\sum_{i=1}^N Y_i(\boldsymbol{T}=\boldsymbol{t}_{2i})-Y_i(\boldsymbol{T}=\boldsymbol{t}_{1i})$.
Because $Y_i(\boldsymbol{T}=\boldsymbol{t}_{2i})$ is unobserved for all $i$, we instead focus our analysis around the following estimand:
\[
\tau=\sum_{i=1}^N \mathbb{E}(Y(\boldsymbol{T}=\boldsymbol{t}_{2i})|\boldsymbol{X}_i)-Y_i(\boldsymbol{T}=\boldsymbol{t}_{1i})
\]
$\tau$ is the TEA. With $Y_i(\boldsymbol{T}=\boldsymbol{t}_{1i})$ observed for all $i$, we only need to estimate $\mathbb{E}(Y(\boldsymbol{T}=\boldsymbol{t}_{2i})|\boldsymbol{X}_i)$ for each $i$ to obtain an estimate of the TEA. Note that this quantity is conditional on $\boldsymbol{X}_i$, formalizing our statement above that, in using the TEA to evaluate a regulation, we are making the assumption that the only factor that would have been different in the no-regulation scenario is the pollutant exposures. Moreover, this causal estimand is unique because it captures the health effects due to changes in multiple continuous pollutants simultaneously. This feature provides an important improvement over traditional regulation health impact analyses, where health impacts are estimated separately for each pollutant. Analyzing each pollutant separately is challenging, as estimates can be biased when pollution exposures are correlated, and may fail to capture synergistic effects.

We now present the assumptions needed to identify $\mathbb{E}(Y(\boldsymbol{T}=\boldsymbol{t}_{2i})|\boldsymbol{X}_i)$ from the observed data. These assumptions are extensions of those of \cite{wu2018matching}.

\vspace{.5cm}

\noindent
\textit{Assumption 1 (A1) Causal Consistency:} $Y_i(\boldsymbol{T}=\boldsymbol{t})=Y_i \text{ if } \boldsymbol{T}_i=\boldsymbol{t}$

\vspace{.5cm}

The causal consistency assumption states that the observed outcome should correspond to the potential outcome under the observed treatment value.

\vspace{.5cm}

\noindent
\textit{Assumption 2 (A2) Weak Unconfoundedness:} Let $I_i(\boldsymbol{t})$ be an indicator function taking value 1 if $\boldsymbol{T}_i=\boldsymbol{t}$ and 0 otherwise. Then weak unconfoundedness is the assumption that $
I_i(\boldsymbol{t}) \independent Y_i(\boldsymbol{T}=\boldsymbol{t})|\boldsymbol{X}_i
$.

\vspace{.5cm}

The weak unconfoundedness assumption was introduced by \cite{imbens2000role} and is commonly used for causal inference with non-binary treatments. In our context, it says that assignment to treatment level $\boldsymbol{t}$ versus any other treatment level is independent of the potential outcome at treatment $\boldsymbol{t}$ conditional on the observed confounders.

\vspace{.5cm}

\noindent
\textit{Assumption 3 (A3) Overlap:} For each $\left\lbrace \boldsymbol{t}_{2i},\boldsymbol{x}_i \right\rbrace$,
$0 < P(\boldsymbol{T}=\boldsymbol{t}_{2i}|\boldsymbol{X}=\boldsymbol{x}_i)$.

\vspace{.5cm}

The overlap assumption in this context differs slightly from the overlap (or positivity) assumption in classic causal inference analyses. It says that, for each unit's set of counterfactual treatment and confounders, the probability of observing that treatment and confounder level together is greater than zero. This ensures that we are not considering counterfactuals outside the space of feasible treatment and confounder combinations.

\vspace{.5cm}

\noindent
\textit{Assumption 4 (A4) Conditional Smoothness:} 
Let $\Theta_{\boldsymbol{t}}=[t_1-\delta_1,t_1+\delta_1] \times \cdots \times [t_Q-\delta_Q,t_Q+\delta_Q]$, $t_q$ represents the q$^{th}$ element of $\boldsymbol{t}$, and $\delta_1,...,\delta_Q$ are positive sequences tending to zero.
Then smoothness is the assumption that
$$\lim_{\delta_1,...,\delta_Q\to 0} \mathbb{E}(Y|\boldsymbol{X},\boldsymbol{T}\in \Theta_{\boldsymbol{t}})=\mathbb{E}(Y|\boldsymbol{X},\boldsymbol{T}=\boldsymbol{t})$$

\vspace{.5cm}

This multidimensional smoothness assumption is needed due to the continuous nature of the multivariate treatments, which means that we will never have $\boldsymbol{T}=\boldsymbol{t}$ exactly and must instead rely on $\boldsymbol{T}$ within a small neighborhood of $\boldsymbol{t}$.

\vspace{.5cm}

\noindent
Using these four assumptions, we show that $\mathbb{E}(Y(\boldsymbol{T}=\boldsymbol{t}_{2i})|\boldsymbol{X}_i)$ can be identified from observed data.
\begin{equation}\label{eq:identify}
    \begin{split}
        \mathbb{E}(Y(\boldsymbol{T}=\boldsymbol{t}_{2i})|\boldsymbol{X}_i) & = \mathbb{E}(Y(\boldsymbol{T}=\boldsymbol{t}_{2i})|\boldsymbol{X}_i, \boldsymbol{T}=\boldsymbol{t}_{2i}) \quad \text{(by A2)} \\
        &=\mathbb{E}(Y|\boldsymbol{X}_i, \boldsymbol{T}=\boldsymbol{t}_{2i})\quad \quad \quad \quad \quad \text{(by A1)} \\
        & = \lim_{\delta_1,...,\delta_Q\to 0}\mathbb{E}(Y|\boldsymbol{X}_i, \boldsymbol{T} \in \Theta_{\boldsymbol{t}_{2i}}) \quad \quad \text{(by A4)}
    \end{split}
\end{equation}
The expectation in the bottom line of equation~\ref{eq:identify} can be estimated from observed data for small, fixed values of $\delta_1,...,\delta_Q$.

In the next two sections, we introduce methods to perform this estimation with the data described in Section~\ref{s:data}. Before doing so, we clarify a few additional assumptions that will be relied upon in order to interpret a TEA estimate as the number of health events avoided due to the regulation-attributable changes in pollutants. These assumptions, which are not common in causal inference analyses, are needed in the universal regulation scenario due to the complete reliance on observed data under the regulation for estimation. First, we must assume that each of the following relationships would be the same in the regulation and no-regulation scenarios: (1) the pollutant-outcome relationships, (2) the confounder-outcome relationships, and (3) the pollutant-confounder relationships. Second, we must assume that no additional confounders would be introduced in the no-regulation scenario.

\subsection{Matching Estimator of the TEA}\label{ss:matching}
In this section, we introduce a causal inference matching procedure to estimate $\mathbb{E}(Y|\boldsymbol{X}_i, \boldsymbol{T} \in \Theta_{\boldsymbol{t}_{2i}})$ (and thereby the TEA). In Section 3 of the Supplementary Materials, we show that the estimator is consistent, and we describe a bootstrapping approach that can be used to compute uncertainties. The idea behind our matching approach is simple: we will find all units with observed pollutant levels approximately equal to $\boldsymbol{t}_{2i}$ and confounder levels approximately equal to $\boldsymbol{X}_i$, and we will take the average observed outcome value across these units (plus a bias correction term) as an estimate of $\mathbb{E}(Y|\boldsymbol{X}_i, \boldsymbol{T} \in \Theta_{\boldsymbol{t}_{2i}})$.

Let $\boldsymbol{\omega}$ be a $Q$-length vector of pre-specified constants and $\nu$ be a pre-specified scalar. For a column vector $\boldsymbol{b}$, $|\boldsymbol{b}|$ denotes the component-wise absolute value and $||\boldsymbol{b}||=(\boldsymbol{b}'\boldsymbol{A}\boldsymbol{b})^{1/2}$, with $\boldsymbol{A}$ a positive semidefinite matrix. In practice, $\boldsymbol{A}$ will be a covariance matrix so that $||\boldsymbol{b}_1-\boldsymbol{b}_2||$ is the Mahalanobis distance. We let $\varphi(i)$ denote the set of indices of the units matched to unit $i$. Then
\[
\varphi(i)=\left\lbrace j \in 1,...,N : |\boldsymbol{t}_{2i}-\boldsymbol{t}_{1j}|\prec \boldsymbol{\omega} \; \wedge \; ||\boldsymbol{X}_i-\boldsymbol{X}_j||<\nu \right\rbrace
\]
In the first condition here, we are carrying out exact matching within some tolerances $\boldsymbol{\omega}$ on the pollution variables, so that the units matched to unit $i$ have observed pollution levels ($\boldsymbol{t}_{1j}$) almost equal to the counterfactual pollution levels for unit $i$ ($\boldsymbol{t}_{2i}$). This ensures that the matched units have observed pollution values within a small hyperrectangle around $\boldsymbol{t}_{2i}$, i.e. $\boldsymbol{T} \in \Theta_{\boldsymbol{t}_{2i}}$. The second condition carries out Mahalanobis distance matching within some tolerance $\nu$ on the confounders, so that all matched units have confounder values approximately equal to the confounder values for unit $i$. We use separate procedures for matching on the pollution and the confounder values so that we can exercise more direct control over the closeness of the matches on pollution values. Literature on choosing calipers in standard matching procedures \citep{lunt2013selecting} and identifying regions of common support in causal inference \citep{king2006dangers} may provide insight into how to specify $\boldsymbol{\omega}$ and $\nu$. All units that meet the above conditions are used as matches for unit $i$, thus each $i$ is allowed to have a different number of matches, $M_i$. Matching is performed with replacement across the $i$. Then, we estimate $\mathbb{E}(Y|\boldsymbol{X}_i, \boldsymbol{T} \in \Theta_{\boldsymbol{t}_{2i}})$ as
\begin{equation}\label{eq:matching}
\hat{\mathbb{E}}(Y|\boldsymbol{X}_i, \boldsymbol{T} \in \Theta_{\boldsymbol{t}_{2i}})=\frac{P_i}{M_i}\sum_{k\in\varphi(i)} Y^*_k
\end{equation}

We also test varieties of this matching estimator with bias corrections following \cite{abadie2011bias}, because bias corrected matching estimators can provide increased robustness. Let $\hat{\mu}(\boldsymbol{t}_{2i},\boldsymbol{X}_i)$ be a regression-predicted value of $Y$ conditional on $\boldsymbol{t}_{2i}$ and $\boldsymbol{X}_i$. Then the bias corrected estimator has the form 
\[
\hat{\mathbb{E}}(Y|\boldsymbol{X}_i, \boldsymbol{T} \in \Theta_{\boldsymbol{t}_{2i}})=\frac{1}{M_i}\sum_{k\in\varphi(i)} 
(Y_k^*P_i+\hat{\mu}(\boldsymbol{t}_{2i},\boldsymbol{X}_i)-\hat{\mu}(\boldsymbol{t}_{2i},\boldsymbol{X}_k))
\]

While asymptotically matches will be available for all units (as a result of the overlap assumption, A3), in practice there will likely be some units for which no suitable matches can be found in the data. Let $s=1,...,S$ index the units for which 1 or more matches was found. Instead of $\tau$, in finite sample settings we will generally estimate
\[
\tau^*=\sum_{s=1}^S \mathbb{E}(Y|\boldsymbol{X}_s, \boldsymbol{T} \in \Theta_{\boldsymbol{t}_{2s}})-Y_s(\boldsymbol{T}=\boldsymbol{t}_{1s})
\]
by plugging in $\hat{\mathbb{E}}(Y|\boldsymbol{X}_s, \boldsymbol{T} \in \Theta_{\boldsymbol{t}_{2s}})$ for $\mathbb{E}(Y|\boldsymbol{X}_s, \boldsymbol{T} \in \Theta_{\boldsymbol{t}_{2s}})$. This discarding of unmatched units is often called ``trimming'' and the remaining $S$ units called the trimmed sample. This trimming is done to avoid using extrapolation to estimate counterfactual outcomes in areas of the treatment/confounder space that are far from the observed data. Such extrapolation can produce results that are highly biased or model-dependent.

\subsection{BART for TEA Estimation}\label{ss:bart}
In this section, we propose an alternative approach to estimation of the TEA relying on machine learning. Machine learning procedures are typically applied in causal inference as a mechanism for imputation of missing counterfactual outcomes. BART \citep{chipman2010bart} is a Bayesian tree ensemble method. Several recent papers have shown its promising performance in causal inference contexts \citep{hill_bayesian_2011,hahn2017bayesian}.

The general form of the BART model for a continuous outcome $Y$ and a vector of predictors $\boldsymbol{X}$ is $    Y=\sum_{j=1}^J g(\boldsymbol{X};\mathcal{T}_j,\mathcal{M}_j)+\epsilon$, where $j=1,...,J$ indexes the trees in the ensemble, $g$ is a function that sorts each unit into one of a set of $m_j$ terminal nodes, associated with mean parameters $\mathcal{M}_j=\{\mu_1,...,\mu_{m_j}\}$, based on a set of decision rules, $\mathcal{T}_j$. $\epsilon\sim N(0,\sigma^2)$ is a random error term. BART is fit using a Bayesian backfitting algorithm. For estimation of average treatment effects (ATE) in a binary treatment setting, BART is fit to all the observed data and used to estimate the potential outcome under treatment and under control for each unit \citep{hill_bayesian_2011}. The average difference in estimated potential outcomes is computed across the sample to obtain an ATE estimate.

We invoke BART similarly to estimate the TEA, inserting the rates as the outcome. For clarity in this section, we use the notation $Y_i^*$ as the observed outcome rate under the factual pollution levels and $\bar{Y}_i^*$ as the missing counterfactual outcome rate under the counterfactual pollution levels. Boldface versions denote the vector of all outcomes. We fit the following BART model to our data: $Y_i^*=\sum_{j=1}^J g(\boldsymbol{t}_{1i},\boldsymbol{X}_i;\mathcal{T}_j,\mathcal{M}_j)+\epsilon_i$.
We collect $H$ posterior samples of the parameters from this BART model, referred to collectively as $\theta$. For a given posterior sample $h$ and for each unit $i$, we collect a sample from the posterior predictive distribution of $\bar{Y}_i^*$, $p(\bar{Y}_{i}^*|\mathbf{Y}^*)=\int p(\bar{Y}_i^*|\mathbf{Y}^*,\theta)p(\theta|\mathbf{Y}^*)d\theta$.
Denote this posterior predictive sample
$\bar{Y}_i^{*(h)}$. We use these to construct a posterior predictive sample of $\tau$ as $\tau^{(h)}=\sum_{i=1}^N (\bar{Y}_i^{*(h)}-Y_i^*)P_i$. We then estimate $\hat{\tau}=\frac{1}{H}\sum_{h=1}^H \tau^{(h)}$, i.e. the posterior mean. To relate this back to the notation of the previous section, note that this is equivalent to estimating $\hat{\mathbb{E}}(Y|\boldsymbol{X}_i, \boldsymbol{T} \in \Theta_{\boldsymbol{t}_{2i}})=\frac{1}{H} \sum_{h=1}^H \bar{Y}_i^{*(h)}$ for all $i$. A 95\% credible interval is formed with the 2.5\% and 97.5\% percentiles of the $\tau^{(h)}$.

Recall that, with the matching estimator, units for which no matches can be found within the pre-specified tolerances are trimmed to avoid extrapolation, so that we estimate $\tau^*$ rather than $\tau$. BART does not automatically identify units for which extrapolation is necessary to estimate counterfactual outcomes. \cite{hill2013assessing} proposed a two-step method to identify units for trimming with BART which could be adapted for use here. In our simulations and analyses, to ensure comparability of the results from BART and matching, we fit the BART model to the entire dataset but, in estimating the TEA, we omit the same set of units that are trimmed with matching.


\section{Simulations}\label{s:sims}

In this section, we generate synthetic data mimicking the structure of our real data and test the following methods for estimating $\tau^*$: our matching procedure with various types of bias correction, our BART procedure, and a simple Poisson regression. In these simulations, we let $N=5,000$ and $Q=2$, with one pollutant simulated to mimic PM$_{2.5}$ (in $\mu g/m^3$) and the other O$_3$ (in parts per million). Specifically, the observed pollutants are generated as follows: $\boldsymbol{t}_{1i} \sim MVN(\boldsymbol{\mu},\boldsymbol{\Sigma})$, $\boldsymbol{\mu}=\left[12.18 \; \; 0.05\right]'$, $\boldsymbol{\Sigma}=\text{diag}(\left[7.99 \; \; 0.0001\right])$. Then, using $\boldsymbol{t}_{1i}$, we generate the counterfactual pollution levels as $\boldsymbol{t}_{2i}=\boldsymbol{t}_{1i}+\boldsymbol{z}_i$, $\boldsymbol{z}_i=\left[z_{1i} \; \; z_{2i}\right]$, $z_{1i} \sim Unif(0,7.08)$ and $z_{2i} \sim Unif(0,0.03)$. The result is that the counterfactual pollutant levels are always larger than the observed pollutant levels, but the magnitudes of the differences vary across units and at appropriate scales for each pollutant.

We use the exposure values to generate five confounders, $X_{1i},...,X_{5i}$. We let $X_{hi}=\boldsymbol{t}_{1i}'\boldsymbol{\alpha}_h+\epsilon_{hi}$, where $\epsilon_{hi} \sim N(0,\sigma^2_h)$. In addition to the exposures and confounders, we also generate four random variables used as ``unobserved'' predictors of $Y$, i.e. we use them to generate $Y$ but do not include them in the estimation procedures (they are not confounders because they are not related to exposure). In our real data analysis, there are likely many unobserved predictors of the health outcomes considered (but hopefully no unobserved confounders), thus it is important to evaluate our methods under such conditions. 

In each simulation, we generate $Y_i \sim Poisson(\lambda_i)$, with $\lambda_i$ a function of the observed exposures, confounders, and predictors. We also generate an expected counterfactual outcome for each unit, $\lambda_i^{cf}$ using the same functional form but substituting the counterfactual exposure values for the observed exposure values. We consider three different functional forms for $\lambda_i$, and we refer to the different structures as S-1, S-2, and S-3.  S-1 uses the most complex form to construct $Y$, involving strong non-linearities and interactions both within and across the exposures and confounders. S-2 is slightly less complex, with exposure-confounder interactions excluded. Finally, in S-3 all relationships are linear. See the model forms and parameter values used in each simulation in Section 5 of the Supplementary Materials. Parameter values are chosen so that the distribution of $Y$ is similar to the distribution of our outcomes in the Medicare data.

With these data we first perform matching to determine which units have suitable matches for their $\boldsymbol{t}_{2i}$ in the data and will therefore contribute to the estimation of $\tau^*$. Within each of the three simulation scenarios, we set $\nu=1.94$, which is approximately the $10^{th}$ percentile of the Mahalanobis distances between all the units in the data, and consider three different specifications of $\boldsymbol{\omega}$, the tolerances for the pollutant matches. We use 10\%, 15\%, and 25\% of the standard deviations of the counterfactual pollutant distributions, resulting in $\boldsymbol{\omega}=\left[0.35 \; \; 0.001 \right]$, $\boldsymbol{\omega}=\left[0.53 \; \; 0.002 \right]$, and $\boldsymbol{\omega}=\left[0.88 \; \; 0.003 \right]$. Following matching, we estimate three variations of the matching estimator. The first is the simple matching estimator given in equation~\ref{eq:matching} (Match 1). The second is this matching estimator plus a bias correction from a Poisson regression with linear terms, i.e. $\text{log}(\lambda_i)=\beta_0+ \left[\boldsymbol{t}_{1i}' \; \; \boldsymbol{X}_i' \right]\boldsymbol{\beta}$ (Match 2). The third is the matching estimator with a bias correction from a Poisson regression in which the forms of the exposures and confounders are correctly specified (Match 3). We also apply BART as described in Section~\ref{ss:bart}, fitting the BART model to all the data but estimating the TEA with only the units retained after matching. We also compare these methods to a simple Poisson regression. We fit a Poisson regression with all exposures and confounders included as linear terms (PR 1)
and a Poisson regression model in which the forms of the exposures and confounders are correctly specified (PR 2). We compare each estimate to the true TEA in the trimmed sample, $\tau^*=\sum_{s=1}^S \lambda_s^{cf}-\lambda_s$.

Table~\ref{tab:simresults} contains the simulation results. It shows the proportion of units retained after trimming, the ratio of the true trimmed sample TEA to the true whole sample TEA, and the percent bias and 95\% confidence/credible interval coverage (in parentheses) for each method in each simulation. Across all simulations, substantial portions of the sample are being trimmed (37\%-58\%). These trimmed units account for a disproportionately high amount of the TEA, as $\tau^*/\tau$ is generally much smaller than $S/N$. This reflects the scenario we would expect in our real data, because the units with the highest counterfactual pollution levels are likely to have the largest effect sizes, yet these units are unlikely to be matched since we may observe few or no pollution exposures as high as their counterfactuals.

\begin{table}[h]
\centering
\caption{Percent absolute bias (95\% confidence/credible interval coverage) of the following methods in estimation of $\tau^*$ in 200 simulations: matching with no bias correction (Match 1), matching with a linear bias correction (Match 2), matching with correctly specified bias correction (Match 3), BART, Poisson regression with linear terms (PR 1), correctly specified Poisson regression (PR 2). $\boldsymbol{\omega}$ refers to matching tolerances on pollutants, i.e. 0.1 means that matches are restricted to be within 0.1 standard deviations for each pollutant. $S/N$ is the proportion of the sample retained after trimming, and $\tau^*/\tau$ is the ratio of the true trimmed sample TEA to the true whole sample TEA.}
\centerline{
\begin{tabular}{p{1.4cm}cccllllll}
  \hline
 & $\boldsymbol{\omega}$ & $S/N$ & $\tau^*/\tau$ & Match 1 & Match 2 & Match 3 & BART & PR 1 & PR 2 \\ 
  \hline
\multirow{3}{1.4cm}{\shortstack[l]{S-1\\ $\tau=60577$}} & 0.1 & 0.42 & 0.22 & 0.23 (0.19) & 0.06 (0.98) & 0.02 (0.99) & 0.05 (0.82) & 0.76 (0.00)  & 0.03 (0.72) \\ 
  & 0.15 & 0.54 & 0.30 & 0.30 (0.00) & 0.14 (0.65) & 0.01 (0.95) & 0.07 (0.54) & 0.62 (0.00)  & 0.02 (0.84) \\ 
  & 0.25 & 0.67 & 0.42 & 0.39 (0.00) & 0.22 (0.10) & 0.04 (0.79) & 0.08 (0.44) & 0.50 (0.00)  & 0.02 (0.88) \\ \hline
\multirow{3}{1.4cm}{\shortstack[l]{S-2\\$\tau=49610$}} & 0.1 & 0.42 & 0.22 & 0.31 (0.05) & 0.08 (0.98) & 0.02 (1.00) & 0.07 (0.71) & 0.85 (0.00) & 0.04 (0.73)  \\ 
  & 0.15 & 0.54 & 0.31 & 0.38 (0.00) & 0.16 (0.62) & 0.02 (0.96) & 0.10 (0.38) & 0.69 (0.00) & 0.02 (0.86)  \\ 
  & 0.25 & 0.67 & 0.43 & 0.48 (0.00) & 0.25 (0.09) & 0.04 (0.78) & 0.11 (0.35) & 0.57 (0.00) & 0.02 (0.88)  \\ \hline
\multirow{3}{1.4cm}{\shortstack[l]{S-3\\$\tau=5321$}} & 0.1 & 0.42 & 0.32 & 0.42 (0.77) & 0.01 (0.98) & - & 0.13 (0.97) & 0.03 (0.90) & - \\ 
  & 0.15 & 0.54 & 0.42 & 0.53 (0.60) & 0.17 (0.87) & - & 0.22 (0.94) & 0.13 (0.84) & - \\ 
  & 0.25 & 0.67 & 0.56 & 0.54 (0.44) & 0.15 (0.90) & - & 0.23 (0.94) & 0.15 (0.89) & - \\ 
   \hline
\end{tabular}
}
\label{tab:simresults}
\end{table}

In all cases either Match 3 or PR 2, the methods with correct model specification, achieves the best performance. However, in general we do not know the correct functional form of the model, therefore a comparison of the other methods is more relevant to real data. Among the remaining methods, we generally see that as $\boldsymbol{\omega}$ increases, the bias of the TEA estimates increase. This is consistent with expectations, as a larger $\boldsymbol{\omega}$ allows matches with more distant exposure values, which can lead to inappropriate extrapolation. Match 1 and PR 1 generally are not reliable, with bias consistently greater than 30\% and poor coverage. Thus, we focus on a comparison of BART and Match 2.

BART has the smallest bias in every simulation within S-1 and S-2. This is consistent with previous research, which has shown BART to accurately capture complex functional forms. However, BART and other tree-based methods often struggle with simple linear relationships. This is reflected in the results of S-3, where Match 2 achieves smaller bias.

The takeaways regarding coverage are more complicated. We see that, in all simulations with $\boldsymbol{\omega}=0.1$ and $\boldsymbol{\omega}=0.15$, Match 2 provides superior coverage to BART. However, as the matching tolerance is increased, leading to greater bias, the coverage of Match 2 declines more rapidly than BART's coverage. This is evidenced by BART's superior coverage in all simulations with $\boldsymbol{\omega}=0.25$. 

These results demonstrate the trade-offs that must be considered when choosing the matching tolerances, or more generally when choosing which units to trim. This issue is particularly salient in this setting because our estimand of interest, the TEA, is a sum rather than an average. When computing a sum, the removal of one unit is likely to impact the estimate more than the removal of the same unit when computing an average. The stricter the tolerances, the more units are dropped from the analysis and, generally, the more distant $\tau^*$ will grow from $\tau$. Therefore, the estimated TEA is less and less likely to reflect the population of interest. In our context where greater pollution exposure is unlikely to have protective effects on health, and where most of the counterfactual pollution levels are greater than the factual ones (i.e. $Y(\boldsymbol{T}=\boldsymbol{t}_{2i})-Y(\boldsymbol{T}=\boldsymbol{t}_{1i})>0$), we anticipate that stricter tolerances and the trimming of more units will lead to underestimation of the TEA. However, stricter tolerances reduce the potential for extrapolation and generally give estimates of $\tau^*$ with lower bias. The selection of these tolerances should be considered carefully in the context of each data application. In the simulated data tolerances of 0.1 and 0.15 standard deviations of each pollutant gave reliable results.

\section{Application}\label{s:app}
We return now to the real data described in Section~\ref{s:data}. Because the purpose of the CAAA was to reduce air pollution, we would expect the counterfactual (no-CAAA) pollution exposures to be higher than factual (with-CAAA) pollution exposures in most zipcodes; however, the CAAA resulted in a complex set of regulations that could have led to decreases in some areas and increases in others. In our data, in 27\% of the zipcodes the factual PM$_{2.5}$ and/or O$_3$ exposure estimate is larger than the corresponding counterfactual estimate, indicating that the CAAA increased exposure to at least one of the pollutants. See Section 5 of the Supplementary Materials for a map showing the locations of these zipcodes. This could reflect real increases in pollution exposures due to the CAAA or it could be partially due to differences in the pollution exposure modeling used to produce the factual and counterfactual estimates. In our primary analysis (PA), to be conservative we include these zipcodes with one or both factual pollutants larger than the corresponding counterfactual, and we also perform a sensitivity analysis (SA) in which they are removed. In the SA, we allow all zipcodes to serve as potential matches but the TEA's sum is only taken across zipcodes with both factual pollutant estimates smaller than the corresponding counterfactual estimate.

Within each of the PA and SA, we conduct separate analyses to estimate the health impacts in years 2000 and 2001 due to CAAA-attributable changes in pollution exposures in the year 2000. Prior to analysis, we remove any zipcodes with missing data or with zero Medicare population. For the year 2000, the sample size following these removals is $N=28,155$ zipcodes (SA: $N=20,432$). All descriptive statistics provided are from the year 2000 data. Figures for 2001 deviate little, if at all, due to year-specific missingness and Medicare population size. For each year's data we apply BART, matching with a linear bias correction, and a Poisson regression with linear terms to estimate the TEA for mortality, cardiovascular disease (CVD) hospitalizations, and dementia hospitalizations. We apply these methods using the rates of each health outcome (yearly event count/population size). 

We first apply matching to the data, with 50 bootstrap replicates and with tolerances $\boldsymbol{\omega}=\left[0.56 \; \; 0.77\right]$ and $\nu=2.12$. As in the simulations the $\boldsymbol{\omega}$ values are 0.1 standard deviations of each counterfactual pollutant distribution, and $\nu$ is approximately the 10$^{th}$ percentile of the Mahalanobis distances between all the confounders in the data. These tolerances lead to trimming of 10,313 zipcodes (SA: 8,936), leaving $S=17,842$ zipcodes (SA: $S=11,496$) for estimation of the TEA. The portion of the full dataset retained to estimate the TEA, $S/N=0.63$ (SA: $S/N=0.56$), is similar to that in the simulations. As described in Sections~\ref{s:methods} and~\ref{s:sims}, we fit the BART and Poisson regression to the entire sample but then only include the units retained after trimming for estimation of the TEA.

The total Medicare population in the zipcodes retained after trimming is 15,573,107 (SA: 10,096,548). Table~\ref{tab:features} shows the means and standard deviations of the Medicare population size, Medicare health outcome rates, exposures and confounders in the full dataset, among the discarded/trimmed zipcodes, and in the retained/untrimmed zipcodes used for estimation for the PA. In Section 5 of the Supplementary Materials, we provide a map showing the locations of the discarded and retained zipcodes after trimming for the PA, and an analog of Table~\ref{tab:features} for the SA. The discarded zipcodes are primarily in the midwest, where Figure~\ref{fig:maps} shows some of the highest counterfactual pollutant levels (this result is consistent with our simulation takeaways in Section~\ref{s:sims}), and in major east and west coast cities. Notably, the discarded zipcodes have larger average population size and higher average pollutant exposures (both factual and counterfactual) than the full dataset. As discussed in Section~\ref{s:sims}, this is likely due to the fact that we have sparse observed data at very high exposure levels and therefore it is difficult to find matches for zipcodes with very high counterfactual exposures (which also tend to be the more highly populous zipcodes). We would also expect that many of the largest health benefits of the CAAA may have come in zipcodes with large populations and whose no-CAAA exposures would have been very high, thus our results are likely to be underestimates of the effects in the whole population.

\begin{table}[h]
\centering
\caption{Average (and standard deviation) of Medicare population size, Medicare health outcome rates, pollutant exposures and confounders in the full dataset, only the discarded/trimmed zipcodes, and the retained/untrimmed zipcodes used for estimation.}
\centerline{
\begin{tabular}{rlll}
  \hline
 & Full Dataset & Discarded Units & Retained Units \\ 
  \hline
  Population size & 1091.76 (1514.94) & 1470.52 (1797.91) & 872.83 (1273.76) \\ 
  Mortality (rate per 1,000) & 52.22 (21.29) & 52.79 (21.8) & 51.89 (20.99) \\ 
  Dementia (rate per 1,000) & 16.69 (14.27) & 17.48 (17.19) & 16.23 (12.25) \\ 
  CVD (rate per 1,000) & 72.35 (31.22) & 70.43 (33.64) & 73.45 (29.68) \\ 
  Factual PM$_{2.5}$ ($\mu g /m^3$) & 10.51 (3.76) & 12.26 (3.34) & 9.5 (3.61) \\ 
  Factual O$_3$ (ppb) & 46.92 (6.15) & 47.22 (6.76) & 46.75 (5.76) \\ 
  Counterfactual PM$_{2.5}$ ($\mu g /m^3$) & 14.5 (5.62) & 18.5 (5.85) & 12.18 (3.92) \\ 
  Counterfactual O$_3$ (ppb) & 53.3 (7.74) & 57.7 (8.33) & 50.76 (6.06) \\ 
  poverty (proportion) & 0.11 (0.1) & 0.12 (0.12) & 0.11 (0.08) \\ 
  popdensity (per mi$^2$) & 1146.31 (4426.27) & 2437.78 (6989.55) & 399.81 (1076.25) \\ 
  housevalue (USD) & 103528.58 (82347.49) & 131038.42 (114351.7) & 87627.39 (49523.45) \\ 
  black (proportion) & 0.08 (0.16) & 0.11 (0.2) & 0.06 (0.13) \\ 
  income (USD) & 40222.76 (16041.13) & 43952.52 (20441.86) & 38066.89 (12322.53) \\ 
  ownhome (proportion) & 0.75 (0.15) & 0.71 (0.2) & 0.78 (0.1) \\ 
  hispanic (proportion) & 0.06 (0.12) & 0.08 (0.16) & 0.05 (0.09) \\ 
  education (proportion) & 0.38 (0.18) & 0.4 (0.2) & 0.38 (0.16) \\ 
   \hline
\end{tabular}
}
\label{tab:features}
\end{table}

\begin{figure}[h]
\centering
\includegraphics[scale=.5]{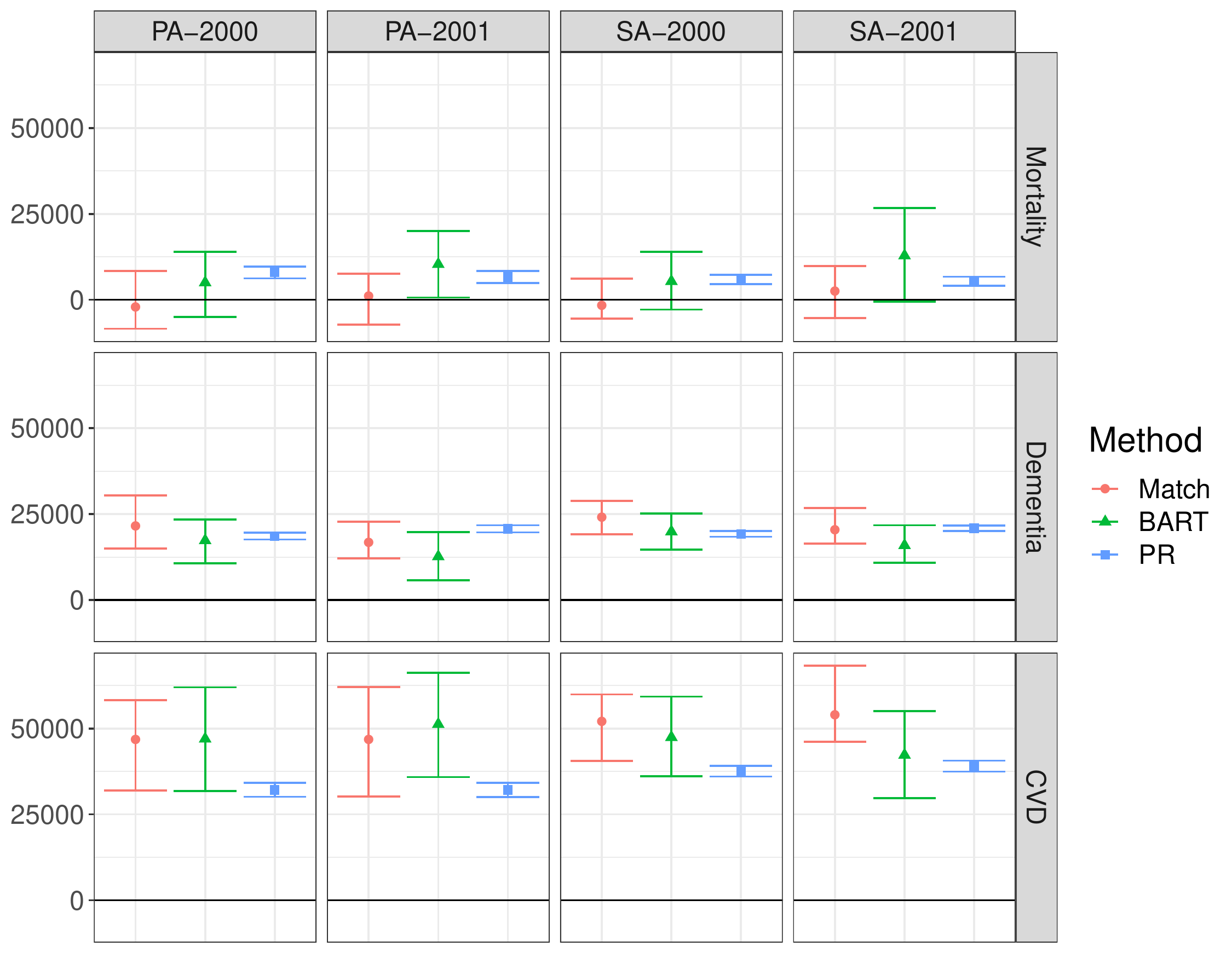}
\caption{Primary analysis (PA) and sensitivity analysis (SA) estimates and 95\% confidence/credible intervals for the TEA in the Medicare population in 2000 and 2001 due CAAA-attributable changes in PM$_{2.5}$ and O$_3$ in the year 2000. Estimation is performed using matching with a linear bias correction, BART, and Poisson regression (PR).}
\label{fig:results}
\end{figure}

The results of both the PA and SA appear in Figure~\ref{fig:results}. We first note that the results of the PA and SA are highly similar, indicating that the zipcodes for which one or both of the factual pollutant exposures are larger than the counterfactual have little impact on the analysis. Matching and BART find limited, although somewhat inconsistent, evidence of effects on mortality. Only the BART analysis for 2001 detects an effect, estimating approximately 10,000 mortalities prevented in the retained zipcodes, with the lower credible interval limit just exceeding zero. For CVD and dementia hospitalizations, all the methods yield large positive point estimates, i.e., large estimates of the number of events avoided in the specified year due to CAAA-attributable pollution changes in 2000. The estimates from BART and matching suggest that approximately 23,000 dementia hospitalizations and approximately 50,000 CVD hospitalizations were avoided in each year. None of the 95\% confidence/credible intervals for these outcomes overlaps zero, providing strong evidence of an effect. The Poisson regression gives all positive and statistically significant TEA estimates, with the significance attributable to extremely narrow confidence intervals. However, the simulation results suggest that these results are likely untrustworthy.

Air pollution literature accumulated over the past 20 years has provided strong evidence to support a causal relationship between long-term PM$_{2.5}$ exposure and mortality \citep{epa2009integrated}. This relationship has also been reported in a recent analysis of Medicare data \citep{di2017air}. One major difference in our study compared to previous cohort studies examining long-term PM$_{2.5}$ exposure and mortality is that we are using zipcode aggregated data while they use individual level data. The use of individual level data allows for adjustments for individual level features, while the use of aggregated data does not. Therefore, results from individual level data are often preferred. Given the complex nature of mortality, our inability to adjust for individual level factors may make mortality rates appear too noisy to detect pollution effects. Our results are consistent with the recent findings of  \cite{henneman2019decreases} who investigate coal emissions exposures and health outcomes at the zipcode level in the Medicare population. They also find limited effects on mortality but strong evidence of harmful effects on CVD hospitalizations.

To demonstrate how the approach introduced in this paper can be used in conjunction with traditional health impact assessments, we provide the EPA's Section 812 Analysis estimates for the number of mortalities and CVD hospitalizations prevented in the year 2000 due to the CAAA in Section 4 of the Supplementary Materials. We also provide additional context to clarify how their estimates can be compared to ours. In short, our approach yields a conservative set of results entirely supported by real, population-level health outcome data. Our results also account for any synergistic effects of the pollutants on health. The traditional approach used in the Section 812 analysis relies on 1) health effect estimates from cohort studies which allow for extensive confounding adjustment but may not fully reflect the population of interest; and 2) a stronger set of modeling assumptions which, while unverifiable, allows for the number of events avoided to be estimated in areas with no support in the real data. Considering the distinct strengths and limitations of our approach and the traditional approach, the results of both types of analysis are likely to provide useful insights about the health impacts of large air pollution regulations.

Finally, we note that in order to interpret our results here as causal, we must be willing to make the strong causal identifying assumptions put forth in Section~\ref{s:methods}. The no unobserved confounding assumption could be called into question, because we have not adjusted for potential behavioral confounders such as smoking and obesity. However, these features may be highly correlated with features we do adjust for, e.g. education. We also make the strong assumption of no interference. It is unclear how a violation of this assumption would impact our results. However, this assumption has been made in most causal inference analyses of air pollution to date \citep{papadogeorgou2018causal,wu2018matching}.

\section{Discussion}\label{s:discuss}
In this paper, we have introduced a causal inference approach for evaluating the health impacts of an air pollution regulation. We developed an estimand called the TEA and proposed a matching and a machine learning method for estimation. Both methods showed promising performance in simulations, particularly in comparison to standard parametric approaches. We implemented these methods to estimate the TEA for mortality, dementia hospitalizations, and CVD hospitalizations in the Medicare population due to CAAA-attributable pollution changes in the year 2000. We found compelling evidence that CAAA-attributable pollution changes prevented large numbers of CVD and dementia hospitalizations. Because more than one third of the zipcodes were trimmed from our analysis to avoid extrapolation, and because the trimmed zipcodes tend to have larger populations and larger improvements in air quality due to the CAAA, we expect that the true number of health events avoided may be considerably larger than our estimates.

While our causal inference approach improves on the traditional approach to regulation evaluation in many ways, there are trade-offs to be considered. In order to avoid extrapolation and strong model-dependence, our methods discard units whose counterfactual pollution/confounder values are far outside the observed pollution/confounder space. This often leads to discarding of units where the highest impacts would be expected. Our estimates tend to have small bias in estimating the effects in the trimmed sample; however the effects in the trimmed sample may be much lower than in the entire original sample. The traditional approach to regulatory evaluation retains all units for estimation, and simply uses parametric models to extrapolate counterfactual outcomes for units with pollutant/confounder values outside areas of support in the observed data. This extrapolation could produce biased estimates, but it is not obvious what the direction of that bias would be. It may be useful to apply and compare both approaches in future regulation evaluations.

The limitations and assumptions of our methods present opportunities for future methodological advancements. In particular, approaches for causal inference with interference could be integrated from recent work \citep{barkley2017causal,Papadogeorgou2019causal}. Because our analyses could be affected by unobserved spatial confounding, future work could amend our matching approach to take into account distance between matched zipcodes \citep{papadogeorgou2018adjusting}. We also note that, because our outcome is rate data, BART's normality assumption may be violated. A BART for count data has been proposed \citep{murray2017log}, but at the time of writing this manuscript, code was not available. Moreover, future work could extend our approach to accommodate individual level data instead of aggregated data. Finally, our approach relies on pollution data from different sources, i.e., factual pollution estimates from hybrid models and counterfactual estimates from the EPA's Section 812 analysis. In the Supplementary Materials, we provide extensive justification for this choice; however, our results may suffer from incompatibility in the pollution exposure estimates, which could be improved upon in future work.

Finally, improvements in pollution exposure data more broadly are essential to increase the reliability of our approach. Statisticians and engineers should work together to produce nationwide factual and counterfactual pollution exposure estimates that are increasingly spatially granular and are more tailored for this type of analysis. These estimates should be accompanied by uncertainties, which could be taken into account in downstream analyses to provide inference that reflects both outcome and exposure uncertainty.

\section{Code}
R code and instructions for reproducing these analyses is available at \url{https://github.com/rachelnethery/AQregulation}.

\section{Disclaimer}
This manuscript has been reviewed by the U.S. Environmental Protection Agency and approved for publication. The views expressed in this manuscript are those of the authors and do not necessarily reflect the views or policies of the U.S. Environmental Protection Agency.

\section{Acknowledgements}
The authors gratefully acknowledge funding from NIH grants 5T32ES007142-35, R01ES024332, R01ES026217, P50MD010428, DP2MD012722, R01ES028033, and R01MD012769; HEI grant 4953-RFA14-3/16-4; and EPA grant 83587201-0.

\section{Supplementary Materials}

\subsection{Traditional Approaches to Health Impact Assessments for Air Pollution Regulations}

Several tools have been developed worldwide that vary in complexity to estimate the potential health implications of changes in air quality including WHO’s AirQ+, Aphekom (Improving Knowledge and Communication for Decision Making on Air Pollution and Health in Europe), and U.S. EPA’s Environmental Benefits Mapping and Analysis Program – Community Edition (BenMAP – CE) \citep{goudarzi2012estimation,pascal2013assessing,sacks2018environmental}. Compared to the other approaches, BenMAP – CE is advantageous because of its flexibility to conduct analyses ranging from local to global in scale \citep{sacks2018environmental}. As a result of this flexibility, BenMAP-CE is able to combine data from simulated air quality models with pollutant-health exposure-response functions (ERF) to estimate the number of health outcomes prevented due to changes in pollutant exposures within a given year. Air quality modeling softwares, like the ECHAM/MESSy Atmospheric Chemistry-Climate Model (EMAC) \citep{jockel2006atmospheric} and the Community Multiscale Air Quality Modeling System (CMAQ) \citep{cmaq}, combine emissions inventories, meteorological data, and atmospheric chemistry models to estimate gridded pollution concentrations across large areas. In regulation evaluations, BenMAP-CE uses output from various modeling applications such as those mentioned above to estimate gridded factual (with-regulation) pollution exposures and counterfactual (without-regulation) pollution exposures for a given year. 

Pollutant-health ERFs are then used to estimate the number of health events prevented due to an improvement in air quality by using a health effect estimate, i.e. a linear model coefficient capturing the relationship between exposure to an air pollutant and the risk of a health event from a peer-reviewed published epidemiologic study, along with additional inputs including the estimated change in a pollutant concentration (from the air quality modeling software), the baseline incidence rate of the health event, and the size of the population exposed. Our explanation of the ERFs follows the one provided in the supporting materials for the most recent Section 812 Analysis, specifically the document entitled Health and Welfare Benefits Analyses to Support the Second Section 812 Benefit-Cost Analysis of the Clean Air Act \citep{epa2011benefits}. The units of analysis in the ERF approach are typically grid cells. For each grid cell separately, the ERF estimates $\Delta y$, the difference in the health outcome of interest under the factual and counterfactual pollutant levels. The ERF requires the following four inputs: 1) an effect estimate from a previous epidemiologic study relating pollutant exposures to the health outcome ($\beta$); 2) a baseline incidence rate for the health event of interest ($\pi_0$); the population size in the grid cell ($P$); and the difference in the counterfactual and factual pollutant estimate for the grid cell ($\Delta x$). These inputs are plugged into the following formula:
\[ \Delta y=\pi_0 P (e^{\beta \Delta x} -1) \]
$\Delta y$ is computed for each grid cell separately, and the results are summarized to obtain the total number of the health event prevented by regulation-attributable changes in the pollutant. This procedure is generally performed for each relevant pollutant/health combination as described below. 

In conducting regulation-based health impact analyses, EPA examines pollutant/health outcome combinations for which the evidence base is sufficient to conclude that a causal or likely-to-be causal relationship exists, as discussed in EPA’s Integrated Science Assessments \citep{epa2015preamble}, and for which there are published epidemiologic studies available that have examined the exposure-response relationship (to obtain the health effect estimate used by BenMAP-CE). With our proposed approach, we do not rely on this threshold of evidence to dictate the health outcomes evaluated, but instead leverage relationships detected in our observed data for estimation and, if the causal identifying assumptions we lay out in the main manuscript are met, then we are assured that we are capturing causal effects of the pollutant changes on the health outcome. Thus, we are able to analyze effects of the regulation on health outcomes for which the current evidence base is not as rich (e.g., dementia) as is available for some health outcomes (e.g., mortality and cardiovascular effects). 

\subsection{Data}
\subsubsection{Pollution Data}
We chose to use factual exposures from the hybrid models instead of the factual exposure estimates produced for the EPA's Section 812 Analysis for two primary reasons. First, these hybrid models have been shown to have excellent predictive performance and are validated for investigating pollution-health relationships at the population level (\citeauthor{di2017air}, \citeyear{di2017air}a, \citeyear{di2017association}b). Second, the hybrid model pollution exposures are estimated at much higher spatial resolution than the EPA's factual exposure estimates, which can be advantageous for quantification of exposure-health relationships. Because our analysis will rely on exposure-health relationships in the factual data to estimate counterfactual outcomes (and thereby causal effects), high resolution factual pollution data are essential.

\subsubsection{Medicare Data}
Here we describe how we constructed counts of each health event from the Medicare data. Mortality counts are created by counting the number of cohort members in each zipcode with dates of death in the range January 1-December 31 of the specified year. In constructing hospitalization counts, we only count each cohort member's first hospitalization (of each type) in the specified year. Cardiovascular hospitalizations are those with a primary diagnosis ICD-9 code 390.xx-459.xx. Dementia hospitalizations are those with either primary or secondary diagnosis Parkinson's Disease (ICD-9: 332.xx), Alzheimer's Disease (ICD-9: 331.0x), or dementia (ICD-9: 290.xx). This definition follows recent work on air pollution and dementia by \cite{kioumourtzoglou2015long}. 

\subsection{Asymptotic Properties of the Matching Estimator and Bootstrapping}

In this section, we discuss the asymptotic properties of our TEA matching estimator, and we provide a bootstrapping approach for obtaining uncertainties for the TEA estimate. We must show that $\hat{\mathbb{E}}(Y|\boldsymbol{X}_i, \boldsymbol{T} \in \Theta_{\boldsymbol{t}_{2i}})\rightarrow \mathbb{E}(Y|\boldsymbol{X}_i, \boldsymbol{T} \in \Theta_{\boldsymbol{t}_{2i}})$ for all $i$ in order to ensure that $\hat{\tau}\rightarrow\tau$ (recall that asymptotically all units have matches so that we are estimating $\tau$ rather than $\tau^*$). Causal inference analyses typically apply matching procedures that identify a common, fixed number of matches ($M$) for each unit. \cite{abadie2006large} showed the consistency of the matching estimator with fixed $M$ and laid out the conditions for $\sqrt{N}$-consistency. However, these results are not applicable to our estimator because we allow for the number of matches to vary across units ($M_i$) and because our treatment variable is multivariate rather than binary.

To show that each $\hat{\mathbb{E}}(Y|\boldsymbol{X}_i, \boldsymbol{T} \in \Theta_{\boldsymbol{t}_{2i}})$ is consistent, we treat $M_i$, the number of matches for unit $i$, as a function of $N$, so that $M_i\rightarrow\infty$ as $N\rightarrow\infty$. This condition indeed holds as long as the overlap assumption above (A3) is met. In the binary treatment case, \cite{heckman1998matching} show that, when $M$ is allowed to go to infinity as $N$ goes to infinity, matching estimators of the average treatment effect on the treated are asymptotically linear and therefore consistent (although not $\sqrt{N}$- consistent). This emerges from the fact that the estimators of the expected counterfactuals can be represented as kernel regression or local polynomial regression, which are asymptotically linear estimators for every set of confounder values, $\boldsymbol{X}_i$. The same result applies to our estimator of $\mathbb{E}(Y|\boldsymbol{X}_i, \boldsymbol{T} \in \Theta_{\boldsymbol{t}_{2i}})$ and its consistency, under the conditions given in \cite{heckman1998matching}, follows directly from these results. However, the convergence rate is not $\sqrt{N}$.

While some relevant asymptotic normality results (in $M$) are also provided by \cite{heckman1998matching}, these may not be reliable in our setting in which $M_i$ is likely to be small for many $i$. Therefore, we prefer to rely on bootstrapping to obtain inference in this setting. It is a well-known result that the bootstrap fails for matching estimators with fixed $M$ \citep{abadie2008failure}, however this failure is precisely due to the fixing of $M$. The bootstrap is valid for kernel regression \citep{hall1992bootstrap} and therefore is also valid for our estimator, which can be represented as a kernel regression. However, given the unique structure of our data in this setting, where each unit serves as both (1) a unit that we seek matches for and (2) as a potential match for all other units, how to carry out the bootstrap in practice is not obvious.

We want to obtain the empirical distribution of $\tau^*$ using the bootstrap. If we implement the naive bootstrap (resample from the trimmed sample with replacement), some units from the trimmed sample will not appear in the bootstrap sample. Therefore, the set of potential matches changes in the bootstrap sample, meaning that units that had 1 or more matches in the original sample may not have any matches in the bootstrap sample. Yet, to obtain the empirical distribution of $\tau^*$, we must ensure that no units are trimmed when estimating $\tau^*$ in the bootstrap sample. To overcome this issue, we propose the following approach to constructing bootstrap confidence intervals for $\tau$, which has provided reliable results in simulation studies.
\begin{enumerate}
    \item Resample $S$ units with replacement from the trimmed sample, let $b_1=1,...,B_1$ index these units. These are the units whose estimated causal effects we will sum to produce the bootstrap estimate of $\tau^*$, i.e. the units for which we will seek matches to estimate their counterfactual outcomes.
    \item Resample $N$ units with replacement from the entire sample, let $b_2=1,...,B_2$ index these units. This set of units will serve only as potential matches for those from step 1.
    \item For each unit $b_1=1,...,B_1$, find matched units $\varphi(b_1)$ in $b_2=1,...,B_2$ using the following approach. Let $\pi(b_1)=\left\lbrace b_2 \in 1,...,B_2 : |\boldsymbol{t}_{2b_1}-\boldsymbol{t}_{1b_2}|\prec \boldsymbol{\omega} \; \wedge \; ||\boldsymbol{X}_{b_1}-\boldsymbol{X}_{b_2}||<\nu \right\rbrace$. Let \\ $\phi (b_1)=\left\lbrace j\in 1,...,B_2: \sum_{b_2 \neq j} I(||\boldsymbol{t}_{2b_1}-\boldsymbol{t}_{1b_2}||<||\boldsymbol{t}_{2b_1}-\boldsymbol{t}_{1j}||) \leq c \right\rbrace$, i.e. the set of indices of the units $b_2$ with the $c$ smallest values of $||\boldsymbol{t}_{2b_1}-\boldsymbol{t}_{1b_2}||$, with $c$ a small, pre-specified integer. Define $\rho(b_1)=\text{argmin}_{k\in \phi(b_1)} ||\boldsymbol{X}_{b_1} -\boldsymbol{X}_{k}|| $, then
    \[
    \varphi(b_1) = \left\{\begin{array}{ll}
        \pi(b_1), & \text{if } \pi(b_1)\neq \varnothing\\
        \rho(b_1), & \text{otherwise}
        \end{array}\right.
    \]
    With this step, we first seek to find matches for $b_1$ among the $b_2=1,...B_2$ using the same matching procedure originally applied. In order to ensure that no units are trimmed, if no matches are found with that procedure then we find the unit $b_2$ with the smallest Mahalanobis distance on confounders among the $c$ units with the smallest Mahalanobis distance on pollutants, and we choose that unit as a single match for $b_1$. 
    \item For $b_1=1,...,B_1$, compute $\hat{\mathbb{E}}(Y|\boldsymbol{X}_{b_1}, \boldsymbol{T} \in \Theta_{\boldsymbol{t}_{2b_1}})=\frac{P_{b_1}}{M_{b_1}}\sum_{k\in\varphi(b_1)} Y^*_k$ (adding a bias correction if one was used to compute the point estimate) and estimate $\tau^*$ as above
\end{enumerate}
Repeat this procedure $B$ times to get bootstrap estimates $\left\lbrace \tau^*_1,...,\tau^*_B\right\rbrace$. Then the bootstrap confidence limits are constructed from the percentiles of this empirical distribution, i.e. the 2.5 and 97.5 percentiles are used to create a 95\% confidence interval for $\tau^*$. 

\subsection{EPA's Section 812 Analysis Results}
Here we provide the estimates of mortalities and CVD hospitalizations prevented from the health impact assessment in the EPA's Section 812 Analysis of the CAAA \citep{epa2011benefits} (the Section 812 Analysis does not investigate dementia hospitalizations), and we discuss important context that should be considered when comparing these results to our results. While these two analyses seek to estimate the same quantities, the number of each health event prevented due to CAAA-attributable changes in PM$_{2.5}$ and O$_3$, they take very different statistical approaches that warrant a formal comparison. The strengths and limitations of each approach are detailed in the final two paragraphs of this section. An important piece of context to keep in mind when directly comparing the results is the different population sizes for which the analyses are conducted, resulting from both the different age groups analyzed and the trimming involved in our method. Due to the different population sizes under study, the estimated number of events from the two methods should not be compared directly but instead compared as a proportion of the underlying population size. We provide the year-2000 population sizes, which we refer to as denominators, for the different analyses.

For the year 2000, the Section 812 Analysis estimates that the CAAA-attributable changes in PM$_{2.5}$ prevented 110,000 mortalities in adults age 30+, and the changes in O$_3$ prevented 1,400 mortalities across all ages (denominator all Americans 30+: 162,603,304). Thus, they estimate that mortality was prevented due to the CAAA in approximately .07\% of the population under study. They also estimate that 26,000 CVD hospitalizations were prevented across all ages in 2000 (denominator all Americans: 281,421,906), i.e., CVD hospitalizations were prevented in approximately 0.009\% of the population under study due to the CAAA.

Our approach finds inconsistent evidence regarding mortalities prevented due to the CAAA in the Medicare population in the zipcodes retained after trimming (denominator: 15,573,107). Matching detects no reductions in 2000 or 2001 and BART estimates approximately 10,000 mortalities prevented for 2001 only. The BART estimate for 2001 suggests that mortalities were prevented in 0.06\% of the population under study due to the CAAA. Both matching and BART estimate that approximately 50,000 CVD hospitalizations were prevented in 2000 (same denominator as above), suggesting that CVD hospitalizations were avoided in 0.3\% of the population under study due to the CAAA. When comparing the CVD estimates to the Section 812 estimates, note that ERF for CVD in the Section 812 analysis is a pooled estimate from studies that use different sets of ICD-9 codes to define CVD, none of which align perfectly with our set of ICD-9 codes for CVD.

Before considering the strengths and limitations of our method and traditional approach used in the Section 812 Analysis, we make note of the different approaches used by the methods to handle areas whose counterfactual exposures and/or confounders fall outside the range of support of observed data. Our approach removes such areas from the analysis, while the traditional approach uses parametric models to extrapolate the health estimates for these areas. Each approach to handling this issue could be considered a strength or a weakness, depending on one's perspective. Through trimming, our method's results rely on fewer modeling assumptions and are more data-supported, although they reflect a more limited population. The traditional approach is able to produce results for a broader population, but due to the strong modeling assumptions and extrapolation needed to do so, it is possible that the results could be biased. Considering these trade-offs, both sets of results provide important insights into the health impacts of regulations.

There are several appealing features of the traditional approach used in the Section 812 analysis that are not shared by our approach. The ERFs rely on
epidemiologic studies such as the American Cancer Society's Cancer Prevention II study \citep{pope2002lung} and the Harvard Six Cities Study \citep{laden2006reduction}, which are built upon individual level data. As discussed in the main manuscript, generally results from individual level data would be preferred to ecologic data, due to increased ability to control for confounding and eliminate noise. Moreover, the Section 812 analysis has the advantage of using factual and counterfactual pollution exposure estimates that were produced in a consistent manner and were designed specifically for that analysis. The spatial coarseness of their exposure estimates is not as problematic in the traditional approach, because they are used as inputs into pre-specified ERFs rather than using them to analyze pollution-health relationships. For reasons discussed in Section 2 of the Supplementary Materials, we believe that the use of the hybrid-model factual pollution estimates is crucial for our analysis; however, our results may suffer from incompatibility in the factual and counterfactual pollution exposure estimates and from the spatial coarseness of the counterfactual estimates. 

While the traditional approach has some advantages, it is also overly simplistic in numerous ways that are improved upon by our approach. First, the epidemiologic studies upon which the ERFs are built may not be representative of the population to which inference is made. Our approach improves on this by using real population-level data from the population under study. Second, the ERFs generally impose a linear relationship between the pollutants and health outcomes. 
Notably, a linear function can lead to extremely high extrapolated estimates of counterfactual outcomes in areas where counterfactual pollution levels are very high (and outside the range of observed pollutant exposures).
Third, the long-term pollution exposure and health studies from which the ERFs are obtained rely on a user-specified model form for confounding adjustment, generally within a parametric model, which may lead to insufficient adjustment for confounding because in practice the true forms of these models are typically not known. Fourth, the traditional approach treats PM$_{2.5}$ and O$_3$ separately and thereby fails to capture potential interactions between them.

\subsection{Additional Tables and Figures}

\begin{table}[ht]
\centering
\caption{Model forms and parameter values for simulations. Notation generally follows the main manuscript. The $k^{th}$ component of $\boldsymbol{t}_{1i}$ is indexed using the notation $\boldsymbol{t}_{1i_{(k)}}$. $\tilde{X}_{1i},...,\tilde{X}_{4i}$ are the four random predictors, with $\tilde{X}_{1i} \sim N(0,1)$, $\tilde{X}_{2i} \sim Exp(1)$, $\tilde{X}_{3i} \sim Unif(0,1)$, and $\tilde{X}_{4i} \sim N(0,6.25)$.}
\begin{tabular}{rp{9cm}}
  \hline
\multirow{9}{*}{S-1, Outcome Model} & $log(\lambda_i)=\beta_0+\beta_1 X_{1i}+\beta_2 X_{2i}+\beta_3 X_{3i}+\beta_4 X_{4i}+\beta_5 X_{5i}+\beta_6 X_{1i}X_{2i}+\beta_7 X^2_{3i}+\beta_8 exp(X_{4i})(1+exp(X_{4i}))^{-1}+\beta_9 t_{1i_{(1)}}+\beta_{10} t_{1i_{(2)}}+\beta_{11} t_{1i_{(1)}}^2+\beta_{12} t_{1i_{(1)}}t_{1i_{(2)}}+\beta_{13} t_{1i_{(1)}}t_{1i_{(2)}}X_{5i}+\beta_{14} t_{1i_{(1)}}t_{1i_{(2)}}X_{4i}+\beta_{15} \tilde{X}_{1i}+\beta_{16} \tilde{X}_{2i}+\beta_{17} \tilde{X}_{3i}+ \beta_{18} \tilde{X}_{4i}$\\[.3cm]
 & $\beta_0=3,
\beta_1=0.001,\; 
\beta_2=-0.024,\;
\beta_3=0.050,\;
\beta_4=-0.037,\;
\beta_5=0.024,\;
\beta_6=0.034,\;
\beta_7=0.023,\;
\beta_8=-0.035,\;
\beta_9=0.008,\;
\beta_{10}=0.100,\;
\beta_{11}=0.002,\;
\beta_{12}=0.030,\;
\beta_{13}=0.005,\;
\beta_{14}=0.030,\;
\beta_{15}=0.050,\;
\beta_{16}=-0.020,\;
\beta_{17}=0.076,\;
\beta_{18}=-0.03$ \\\hline
\multirow{9}{*}{S-2, Outcome Model} & $log(\lambda_i)=\beta_0+\beta_1 X_{1i}+\beta_2 X_{2i}+\beta_3 X_{3i}+\beta_4 X_{4i}+\beta_5 X_{5i}+\beta_6 X_{1i}X_{2i}+\beta_7 X^2_{3i}+\beta_8 exp(X_{4i})(1+exp(X_{4i}))^{-1}+\beta_9 t_{1i_{(1)}}+\beta_{10} t_{1i_{(2)}}+\beta_{11} t_{1i_{(1)}}^2+\beta_{12} t_{1i_{(1)}}t_{1i_{(2)}}+\beta_{13} \tilde{X}_{1i}+\beta_{14} \tilde{X}_{2i}+\beta_{15} \tilde{X}_{3i}+ \beta_{16} \tilde{X}_{4i}$\\[.3cm]
& $\beta_0=3,
\beta_1=0.001,\; 
\beta_2=-0.024,\;
\beta_3=0.050,\;
\beta_4=-0.037,\;
\beta_5=0.024,\;
\beta_6=0.034,\;
\beta_7=0.023,\;
\beta_8=-0.035,\;
\beta_9=0.008,\;
\beta_{10}=0.100,\;
\beta_{11}=0.002,\;
\beta_{12}=0.030,\;
\beta_{13}=0.050,\;
\beta_{14}=-0.020,\;
\beta_{15}=0.076,\;
\beta_{16}=-0.03$ \\\hline
\multirow{5}{*}{S-3, Outcome Model} & $log(\lambda_i)=\beta_0+\beta_1 X_{1i}+\beta_2 X_{2i}+\beta_3 X_{3i}+\beta_4 X_{4i}+\beta_5 X_{5i}+\beta_6 t_{1i_{(1)}}+\beta_7 t_{1i_{(2)}}+\beta_8 \tilde{X}_{1i}+\beta_9 \tilde{X}_{2i}+\beta_{10} \tilde{X}_{3i}+ \beta_{11} \tilde{X}_{4i}$\\[.3cm]
& $\beta_0=3.5,
\beta_1=0.001,\; 
\beta_2=-0.024,\;
\beta_3=0.050,\;
\beta_4=-0.037,\;
\beta_5=0.024,\;
\beta_6=0.008,\;
\beta_7=0.100,
\beta_8=0.050,\;
\beta_9=-0.020,\;
\beta_{10}=0.076,\;
\beta_{11}=-0.03$\\\hline
\multirow{4}{*}{Confounder Models} & $X_{hi}=\boldsymbol{t}_{1i}'\boldsymbol{\alpha}_h+\epsilon_{hi}$\\[.3cm]
& $\boldsymbol{\alpha}_1= \left[ -0.18 \; \; 12 \right]'$, $\boldsymbol{\alpha}_2= \left[ 0.10 \; \; -5 \right]'$, $\boldsymbol{\alpha}_3= \left[ 0.04 \; \; -3.5 \right]'$,
$\boldsymbol{\alpha}_4=\left[ 0.17 \; \; 7 \right]'$,
$\boldsymbol{\alpha}_5=\left[ 0.04 \; \; -2 \right]'$, 
$\sigma^2_1= 0.09$,
$\sigma^2_2=1.04$,
$\sigma^2_3=10.56$,
$\sigma^2_4=4.12$, 
$\sigma^2_5=5.42$ \\
   \hline
\end{tabular}
\label{tab:simstruct}
\end{table}

\begin{figure}[h!]
\centering
\includegraphics[scale=.38]{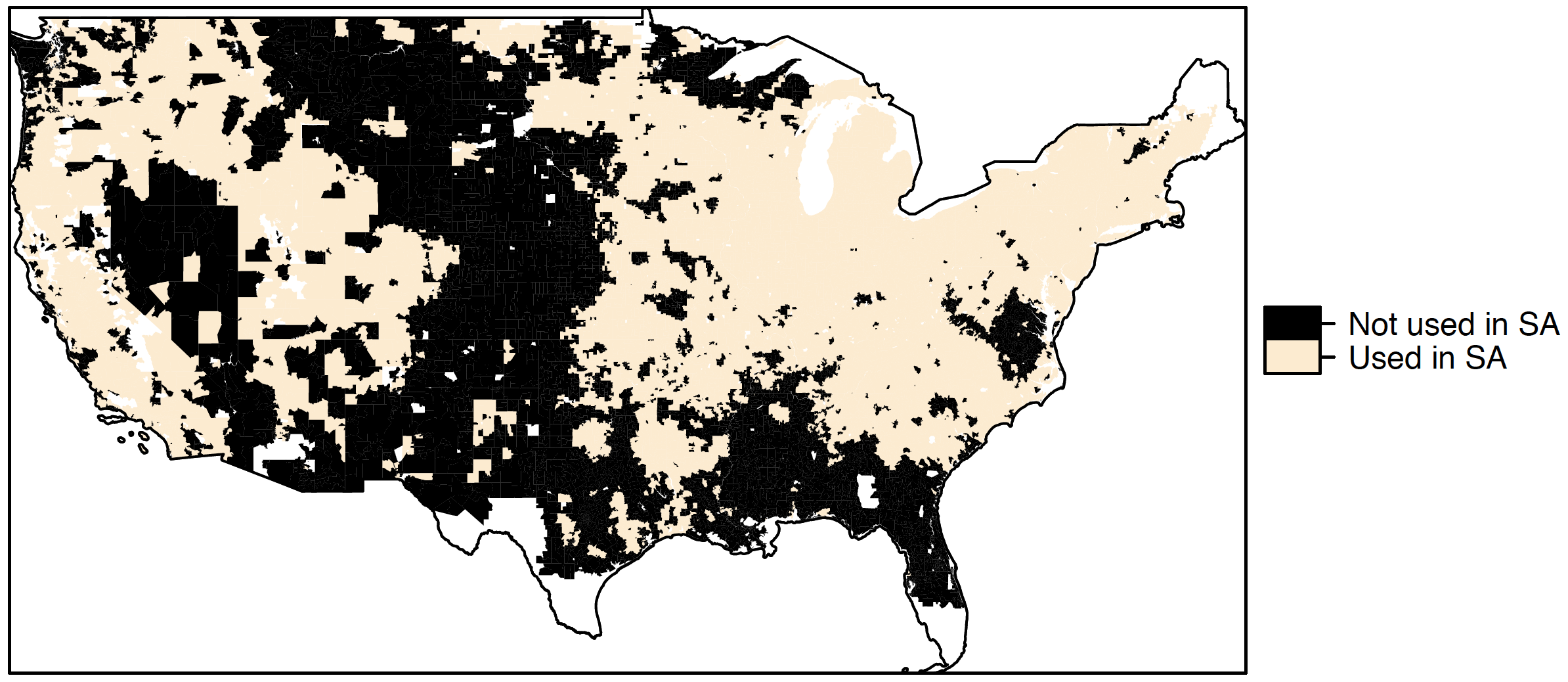}
\caption{Map of the zipcodes used/not used in the sensitivity analysis. Unused zipcodes are those with one or both factual (with-CAAA) pollution estimates larger than the corresponding counterfactual (no-CAAA).}
\end{figure}

\begin{figure}[h!]
\centering
\includegraphics[scale=.38]{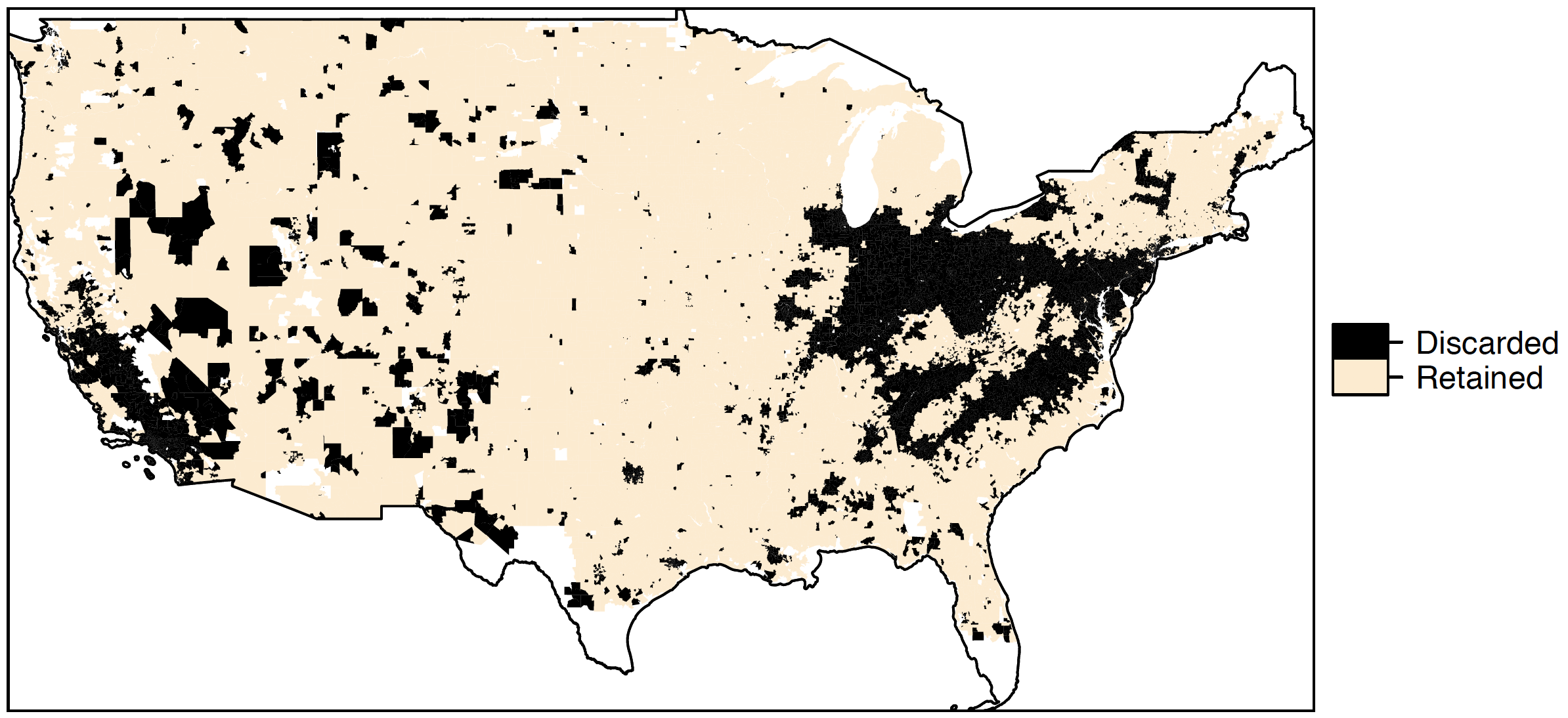}
\caption{Map of the zipcodes retained and discarded due to trimming in the primary analysis.}
\label{fig:trim}
\end{figure}

\begin{table}[ht]
\centering
\caption{Average (and standard deviation) of Medicare population size, Medicare health outcome rates, pollutant exposures and confounders in the sensitivity analysis sample, only the discarded/trimmed zipcodes, and the retained/untrimmed zipcodes used for estimation (year 2000 data).}
\begin{tabular}{rlll}
  \hline
 & SA Sample & Discarded Units & Retained Units \\ 
  \hline
  Population size & 1173.69 (1566.55) & 1553.74 (1818.68) & 878.27 (1261.46) \\ 
  Mortality (rate per 1,000) & 52.29 (20.29) & 52.88 (20.59) & 51.83 (20.05) \\ 
  Dementia (rate per 1,000) & 16.59 (13.46) & 17.54 (15.78) & 15.86 (11.28) \\ 
  CVD (rate per 1,000) & 71.82 (30.34) & 70.61 (32.51) & 72.76 (28.5) \\ 
  Factual PM$_{2.5}$ ($\mu g /m^3$) & 10.57 (3.7) & 12.32 (3.17) & 9.2 (3.49) \\ 
  Factual O$_3$ (ppb) & 46.12 (5.99) & 46.96 (6.65) & 45.46 (5.32) \\ 
  Counterfactual PM$_{2.5}$ ($\mu g /m^3$) & 15.87 (5.45) & 19.39 (5.4) & 13.13 (3.59) \\ 
  Counterfactual O$_3$ (ppb) & 55.03 (7.15) & 58.69 (7.38) & 52.19 (5.48) \\ 
  poverty (proportion) & 0.1 (0.09) & 0.11 (0.11) & 0.1 (0.07) \\ 
  popdensity (per mi$^2$) & 1369.87 (5002.14) & 2591.26 (7278.89) & 420.47 (1108.41) \\ 
  housevalue (USD) & 111633.01 (86652.35) & 133527.37 (113695.03) & 94614.22 (51338.51) \\ 
  black (proportion) & 0.06 (0.15) & 0.09 (0.18) & 0.04 (0.11) \\ 
  income (USD) & 42125.79 (16947.78) & 45068.84 (20720.92) & 39838.12 (12837.77) \\ 
  ownhome (proportion) & 0.75 (0.15) & 0.71 (0.19) & 0.78 (0.1) \\ 
  hispanic (proportion) & 0.05 (0.11) & 0.08 (0.15) & 0.03 (0.07) \\ 
  education (proportion) & 0.38 (0.17) & 0.39 (0.19) & 0.37 (0.16) \\
   \hline
\end{tabular}
\end{table}

\clearpage

\bibliographystyle{chicago}
\bibliography{refs}

\begin{thebibliography}{}

\bibitem[\protect\citeauthoryear{Abadie and Imbens}{Abadie and
  Imbens}{2011}]{abadie2011bias}
Abadie, A. and G.~Imbens (2011).
\newblock Bias-corrected matching estimators for average treatment effects.
\newblock {\em Journal of Business \& Economic Statistics\/}~{\em 29\/}(1),
  1--11.

\bibitem[\protect\citeauthoryear{Abadie and Imbens}{Abadie and
  Imbens}{2006}]{abadie2006large}
Abadie, A. and G.~W. Imbens (2006).
\newblock Large sample properties of matching estimators for average treatment
  effects.
\newblock {\em Econometrica\/}~{\em 74\/}(1), 235--267.

\bibitem[\protect\citeauthoryear{Abadie and Imbens}{Abadie and
  Imbens}{2008}]{abadie2008failure}
Abadie, A. and G.~W. Imbens (2008).
\newblock On the failure of the bootstrap for matching estimators.
\newblock {\em Econometrica\/}~{\em 76\/}(6), 1537--1557.

\bibitem[\protect\citeauthoryear{Abu~Awad, Di, Wang, Choirat, Coull, Zanobetti,
  and Schwartz}{Abu~Awad et~al.}{2019}]{awad2019change}
Abu~Awad, Y., Q.~Di, Y.~Wang, C.~Choirat, B.~Coull, A.~Zanobetti, and
  J.~Schwartz (2019).
\newblock Change in {PM}$_{2.5}$ exposure and mortality among {M}edicare
  recipients.
\newblock {\em In press, Environmental Epidemiology\/}.

\bibitem[\protect\citeauthoryear{Barkley, Hudgens, Clemens, Ali, and
  Emch}{Barkley et~al.}{2017}]{barkley2017causal}
Barkley, B.~G., M.~G. Hudgens, J.~D. Clemens, M.~Ali, and M.~E. Emch (2017).
\newblock Causal inference from observational studies with clustered
  interference.
\newblock {\em arXiv preprint arXiv:1711.04834\/}.

\bibitem[\protect\citeauthoryear{Brook, Rajagopalan, Pope~III, Brook,
  Bhatnagar, Diez-Roux, Holguin, Hong, Luepker, Mittleman, et~al.}{Brook
  et~al.}{2010}]{brook2010particulate}
Brook, R.~D., S.~Rajagopalan, C.~A. Pope~III, J.~R. Brook, A.~Bhatnagar, A.~V.
  Diez-Roux, F.~Holguin, Y.~Hong, R.~V. Luepker, M.~A. Mittleman, et~al.
  (2010).
\newblock {Particulate matter air pollution and cardiovascular disease: an
  update to the scientific statement from the American Heart Association}.
\newblock {\em Circulation\/}~{\em 121\/}(21), 2331--2378.

\bibitem[\protect\citeauthoryear{Chipman, George, McCulloch, et~al.}{Chipman
  et~al.}{2010}]{chipman2010bart}
Chipman, H.~A., E.~I. George, R.~E. McCulloch, et~al. (2010).
\newblock {BART}: {B}ayesian additive regression trees.
\newblock {\em The Annals of Applied Statistics\/}~{\em 4\/}(1), 266--298.

\bibitem[\protect\citeauthoryear{Di, Dai, Wang, Zanobetti, Choirat, Schwartz,
  and Dominici}{Di et~al.}{2017}]{di2017association}
Di, Q., L.~Dai, Y.~Wang, A.~Zanobetti, C.~Choirat, J.~D. Schwartz, and
  F.~Dominici (2017).
\newblock Association of short-term exposure to air pollution with mortality in
  older adults.
\newblock {\em Jama\/}~{\em 318\/}(24), 2446--2456.

\bibitem[\protect\citeauthoryear{Di, Rowland, Koutrakis, and Schwartz}{Di
  et~al.}{2017}]{di2017hybrid}
Di, Q., S.~Rowland, P.~Koutrakis, and J.~Schwartz (2017).
\newblock A hybrid model for spatially and temporally resolved ozone exposures
  in the continental united states.
\newblock {\em Journal of the Air \& Waste Management Association\/}~{\em
  67\/}(1), 39--52.

\bibitem[\protect\citeauthoryear{Di, Wang, Zanobetti, Wang, Koutrakis, Choirat,
  Dominici, and Schwartz}{Di et~al.}{2017}]{di2017air}
Di, Q., Y.~Wang, A.~Zanobetti, Y.~Wang, P.~Koutrakis, C.~Choirat, F.~Dominici,
  and J.~D. Schwartz (2017).
\newblock Air pollution and mortality in the {M}edicare population.
\newblock {\em New England Journal of Medicine\/}~{\em 376\/}(26), 2513--2522.

\bibitem[\protect\citeauthoryear{Goudarzi, Mohammadi, Ahmadi~Angali, Neisi,
  Babaei, Mohammadi, Soleimani, and Geravandi}{Goudarzi
  et~al.}{2012}]{goudarzi2012estimation}
Goudarzi, G., M.~Mohammadi, K.~Ahmadi~Angali, A.~Neisi, A.~Babaei,
  B.~Mohammadi, Z.~Soleimani, and S.~Geravandi (2012).
\newblock Estimation of health effects attributed to {NO2} exposure using
  {AirQ} model.
\newblock {\em Archives of Hygiene Sciences\/}~{\em 1\/}(2), 59--66.

\bibitem[\protect\citeauthoryear{Hahn, Murray, and Carvalho}{Hahn
  et~al.}{2017}]{hahn2017bayesian}
Hahn, P.~R., J.~Murray, and C.~M. Carvalho (2017).
\newblock Bayesian regression tree models for causal inference: regularization,
  confounding, and heterogeneous effects.
\newblock {\em arXiv preprint arXiv:1706.09523\/}.

\bibitem[\protect\citeauthoryear{Hall}{Hall}{1992}]{hall1992bootstrap}
Hall, P. (1992).
\newblock On bootstrap confidence intervals in nonparametric regression.
\newblock {\em The Annals of Statistics\/}~{\em 20\/}(2), 695--711.

\bibitem[\protect\citeauthoryear{Heckman, Ichimura, and Todd}{Heckman
  et~al.}{1998}]{heckman1998matching}
Heckman, J., H.~Ichimura, and P.~Todd (1998).
\newblock Matching as an econometric evaluation estimator.
\newblock {\em The Review of Economic Studies\/}~{\em 65\/}(2), 261--294.

\bibitem[\protect\citeauthoryear{Henneman, Choirat, and Zigler}{Henneman
  et~al.}{2019}]{henneman2019decreases}
Henneman, L.~R., C.~Choirat, and C.~M. Zigler (2019).
\newblock Accountability assessment of health improvements in the {U}nited
  {S}tates associated with reduced coal emissions between 2005 and 2012.
\newblock {\em Epidemiology\/}~{\em 30\/}(4), 477–485.

\bibitem[\protect\citeauthoryear{Henneman, Liu, Chang, Mulholland, Tolbert, and
  Russell}{Henneman et~al.}{2019}]{henneman2019air}
Henneman, L.~R., C.~Liu, H.~Chang, J.~Mulholland, P.~Tolbert, and A.~Russell
  (2019).
\newblock Air quality accountability: {D}eveloping long-term daily time series
  of pollutant changes and uncertainties in {A}tlanta, {G}eorgia resulting from
  the 1990 {C}lean {A}ir {A}ct {A}mendments.
\newblock {\em Environment International\/}~{\em 123}, 522--534.

\bibitem[\protect\citeauthoryear{Hill and Su}{Hill and
  Su}{2013}]{hill2013assessing}
Hill, J. and Y.-S. Su (2013).
\newblock Assessing lack of common support in causal inference using bayesian
  nonparametrics: Implications for evaluating the effect of breastfeeding on
  children's cognitive outcomes.
\newblock {\em The Annals of Applied Statistics\/}, 1386--1420.

\bibitem[\protect\citeauthoryear{Hill}{Hill}{2011}]{hill_bayesian_2011}
Hill, J.~L. (2011).
\newblock Bayesian nonparametric modeling for causal inference.
\newblock {\em Journal of Computational and Graphical Statistics\/}~{\em
  20\/}(1), 217--240.

\bibitem[\protect\citeauthoryear{Ho, Imai, King, and Stuart}{Ho
  et~al.}{2007}]{ho2007matching}
Ho, D.~E., K.~Imai, G.~King, and E.~A. Stuart (2007).
\newblock Matching as nonparametric preprocessing for reducing model dependence
  in parametric causal inference.
\newblock {\em Political Analysis\/}~{\em 15\/}(3), 199--236.

\bibitem[\protect\citeauthoryear{Imbens}{Imbens}{2000}]{imbens2000role}
Imbens, G.~W. (2000).
\newblock The role of the propensity score in estimating dose-response
  functions.
\newblock {\em Biometrika\/}~{\em 87\/}(3), 706--710.

\bibitem[\protect\citeauthoryear{Jöckel, Tost, Pozzer, Brühl, Buchholz,
  Ganzeveld, Hoor, Kerkweg, Lawrence, Sander, and Steil}{Jöckel
  et~al.}{2006}]{jockel2006atmospheric}
Jöckel, P., H.~Tost, A.~Pozzer, C.~Brühl, J.~Buchholz, L.~Ganzeveld, P.~Hoor,
  A.~Kerkweg, M.~Lawrence, R.~Sander, and B.~Steil (2006).
\newblock The atmospheric chemistry general circulation model {ECHAM5/MESSy1}:
  consistent simulation of ozone from the surface to the mesosphere.
\newblock {\em Atmospheric Chemistry and Physics Discussions\/}~{\em 6\/}(4),
  6957--7050.

\bibitem[\protect\citeauthoryear{King and Zeng}{King and
  Zeng}{2006}]{king2006dangers}
King, G. and L.~Zeng (2006).
\newblock The dangers of extreme counterfactuals.
\newblock {\em Political Analysis\/}~{\em 14\/}(2), 131--159.

\bibitem[\protect\citeauthoryear{Kioumourtzoglou, Schwartz, Weisskopf, Melly,
  Wang, Dominici, and Zanobetti}{Kioumourtzoglou
  et~al.}{2015}]{kioumourtzoglou2015long}
Kioumourtzoglou, M.-A., J.~D. Schwartz, M.~G. Weisskopf, S.~J. Melly, Y.~Wang,
  F.~Dominici, and A.~Zanobetti (2015).
\newblock Long-term {PM2. 5} exposure and neurological hospital admissions in
  the northeastern {U}nited {S}tates.
\newblock {\em Environmental health perspectives\/}~{\em 124\/}(1), 23--29.

\bibitem[\protect\citeauthoryear{Kreif, Grieve, D{\'\i}az, and Harrison}{Kreif
  et~al.}{2015}]{kreif2015evaluation}
Kreif, N., R.~Grieve, I.~D{\'\i}az, and D.~Harrison (2015).
\newblock Evaluation of the effect of a continuous treatment: A machine
  learning approach with an application to treatment for traumatic brain
  injury.
\newblock {\em Health Economics\/}~{\em 24\/}(9), 1213--1228.

\bibitem[\protect\citeauthoryear{Laden, Schwartz, Speizer, and Dockery}{Laden
  et~al.}{2006}]{laden2006reduction}
Laden, F., J.~Schwartz, F.~E. Speizer, and D.~W. Dockery (2006).
\newblock Reduction in fine particulate air pollution and mortality: extended
  follow-up of the {H}arvard {S}ix {C}ities study.
\newblock {\em American Journal of Respiratory and Critical Care
  Medicine\/}~{\em 173\/}(6), 667--672.

\bibitem[\protect\citeauthoryear{Louizos, Shalit, Mooij, Sontag, Zemel, and
  Welling}{Louizos et~al.}{2017}]{louizos2017causal}
Louizos, C., U.~Shalit, J.~M. Mooij, D.~Sontag, R.~Zemel, and M.~Welling
  (2017).
\newblock Causal effect inference with deep latent-variable models.
\newblock In {\em Advances in Neural Information Processing Systems}, pp.\
  6446--6456.

\bibitem[\protect\citeauthoryear{Lunt}{Lunt}{2013}]{lunt2013selecting}
Lunt, M. (2013).
\newblock Selecting an appropriate caliper can be essential for achieving good
  balance with propensity score matching.
\newblock {\em American Journal of Epidemiology\/}~{\em 179\/}(2), 226--235.

\bibitem[\protect\citeauthoryear{Moolgavkar}{Moolgavkar}{2003}]{moolgavkar2003air}
Moolgavkar, S. (2003).
\newblock Air pollution and daily deaths and hospital admissions in {L}os
  {A}ngeles and {C}ook counties.
\newblock {\em Revised analyses of time-series studies of air pollution and
  health. Special report. Boston, MA: Health Effects Institute\/}, 183--198.

\bibitem[\protect\citeauthoryear{Murray}{Murray}{2017}]{murray2017log}
Murray, J.~S. (2017).
\newblock Log-linear {B}ayesian additive regression trees for categorical and
  count responses.
\newblock {\em arXiv preprint arXiv:1701.01503\/}.

\bibitem[\protect\citeauthoryear{Nethery and Dominici}{Nethery and
  Dominici}{2019}]{nethery2019estimating}
Nethery, R.~C. and F.~Dominici (2019).
\newblock Estimating pollution-attributable mortality at the regional and
  global scales: challenges in uncertainty estimation and causal inference.
\newblock {\em European Heart Journal\/}.

\bibitem[\protect\citeauthoryear{Papadogeorgou, Choirat, and
  Zigler}{Papadogeorgou et~al.}{2018}]{papadogeorgou2018adjusting}
Papadogeorgou, G., C.~Choirat, and C.~M. Zigler (2018).
\newblock Adjusting for unmeasured spatial confounding with distance adjusted
  propensity score matching.
\newblock {\em Biostatistics\/}~{\em 20\/}(2), 256--272.

\bibitem[\protect\citeauthoryear{Papadogeorgou and Dominici}{Papadogeorgou and
  Dominici}{2018}]{papadogeorgou2018causal}
Papadogeorgou, G. and F.~Dominici (2018).
\newblock A causal exposure response function with local adjustment for
  confounding: A study of the health effects of long-term exposure to low
  levels of fine particulate matter.
\newblock {\em arXiv preprint arXiv:1806.00928\/}.

\bibitem[\protect\citeauthoryear{Papadogeorgou, Mealli, and
  Zigler}{Papadogeorgou et~al.}{2019}]{Papadogeorgou2019causal}
Papadogeorgou, G., F.~Mealli, and C.~M. Zigler (2019).
\newblock Causal inference with interfering units for cluster and population
  level treatment allocation programs.
\newblock {\em In press, Biometrics\/}.

\bibitem[\protect\citeauthoryear{Pascal, Corso, Chanel, Declercq, Badaloni,
  Cesaroni, Henschel, Meister, Haluza, Martin-Olmedo, et~al.}{Pascal
  et~al.}{2013}]{pascal2013assessing}
Pascal, M., M.~Corso, O.~Chanel, C.~Declercq, C.~Badaloni, G.~Cesaroni,
  S.~Henschel, K.~Meister, D.~Haluza, P.~Martin-Olmedo, et~al. (2013).
\newblock Assessing the public health impacts of urban air pollution in 25
  {E}uropean cities: results of the {A}phekom project.
\newblock {\em Science of the Total Environment\/}~{\em 449}, 390--400.

\bibitem[\protect\citeauthoryear{Pope~III, Burnett, Thun, Calle, Krewski, Ito,
  and Thurston}{Pope~III et~al.}{2002}]{pope2002lung}
Pope~III, C.~A., R.~T. Burnett, M.~J. Thun, E.~E. Calle, D.~Krewski, K.~Ito,
  and G.~D. Thurston (2002).
\newblock Lung cancer, cardiopulmonary mortality, and long-term exposure to
  fine particulate air pollution.
\newblock {\em Jama\/}~{\em 287\/}(9), 1132--1141.

\bibitem[\protect\citeauthoryear{Power, Adar, Yanosky, and Weuve}{Power
  et~al.}{2016}]{power2016exposure}
Power, M.~C., S.~D. Adar, J.~D. Yanosky, and J.~Weuve (2016).
\newblock Exposure to air pollution as a potential contributor to cognitive
  function, cognitive decline, brain imaging, and dementia: a systematic review
  of epidemiologic research.
\newblock {\em Neurotoxicology\/}~{\em 56}, 235--253.

\bibitem[\protect\citeauthoryear{Rubin}{Rubin}{1974}]{rubin1974estimating}
Rubin, D.~B. (1974).
\newblock Estimating causal effects of treatments in randomized and
  nonrandomized studies.
\newblock {\em Journal of Educational Psychology\/}~{\em 66\/}(5), 688.

\bibitem[\protect\citeauthoryear{Rubin}{Rubin}{1980}]{rubin1980randomization}
Rubin, D.~B. (1980).
\newblock Randomization analysis of experimental data: {T}he {F}isher
  randomization test comment.
\newblock {\em Journal of the American Statistical Association\/}~{\em
  75\/}(371), 591--593.

\bibitem[\protect\citeauthoryear{Sacks, Lloyd, Zhu, Anderton, Jang, Hubbell,
  and Fann}{Sacks et~al.}{2018}]{sacks2018environmental}
Sacks, J.~D., J.~M. Lloyd, Y.~Zhu, J.~Anderton, C.~J. Jang, B.~Hubbell, and
  N.~Fann (2018).
\newblock The {E}nvironmental {B}enefits {M}apping and {A}nalysis
  {P}rogram--{C}ommunity {E}dition {(BenMAP--CE)}: {A} tool to estimate the
  health and economic benefits of reducing air pollution.
\newblock {\em Environmental Modelling \& Software\/}~{\em 104}, 118--129.

\bibitem[\protect\citeauthoryear{Stuart}{Stuart}{2010}]{stuart2010matching}
Stuart, E.~A. (2010).
\newblock Matching methods for causal inference: A review and a look forward.
\newblock {\em Statistical Science: A Review Journal of the Institute of
  Mathematical Statistics\/}~{\em 25\/}(1), 1--21.

\bibitem[\protect\citeauthoryear{{US EPA}}{{US EPA}}{2009}]{epa2009integrated}
{US EPA} (2009).
\newblock {Integrated Science Assessment (ISA) For Particulate Matter (Final
  Report, Dec 2009). U.S. Environmental Protection Agency, Washington, DC,
  EPA/600/R-08/139F}.
\newblock Accessed Online: 2019-08-17.

\bibitem[\protect\citeauthoryear{{US EPA}}{{US EPA}}{2011}]{epa2011benefits}
{US EPA} (2011).
\newblock {Benefits and Costs of the Clean Air Act 1990-2020, the Second
  Prospective Study}.
\newblock Accessed Online: 2019-05-08.

\bibitem[\protect\citeauthoryear{{US EPA}}{{US EPA}}{2015}]{epa2015preamble}
{US EPA} (2015).
\newblock {Preamble to the Integrated Science Assessments (ISA). U.S.
  Environmental Protection Agency, Washington, DC, EPA/600/R-15/067}.

\bibitem[\protect\citeauthoryear{{US EPA}}{{US EPA}}{2019}]{cmaq}
{US EPA} (2019).
\newblock {Community Multiscale Air Quality Modeling System (CMAQ)}.
\newblock \url{doi:10.5281/zenodo.107987}.
\newblock Accessed: 2019-05-08.

\bibitem[\protect\citeauthoryear{van Donkelaar, Martin, Li, and Burnett}{van
  Donkelaar et~al.}{2019}]{van2019regional}
van Donkelaar, A., R.~V. Martin, C.~Li, and R.~T. Burnett (2019).
\newblock Regional estimates of chemical composition of fine particulate matter
  using a combined geoscience-statistical method with information from
  satellites, models, and monitors.
\newblock {\em Environmental Science \& Technology\/}.

\bibitem[\protect\citeauthoryear{van Erp, O'Keefe, Cohen, and Warren}{van Erp
  et~al.}{2008}]{van2008evaluating}
van Erp, A.~M., R.~O'Keefe, A.~J. Cohen, and J.~Warren (2008).
\newblock Evaluating the effectiveness of air quality interventions.
\newblock {\em Journal of Toxicology and Environmental Health, Part A\/}~{\em
  71\/}(9-10), 583--587.

\bibitem[\protect\citeauthoryear{{WHO}}{{WHO}}{2019}]{who2019airq}
{WHO} (2019).
\newblock {AirQ+: software tool for health risk assessment of air pollution}.
\newblock Accessed Online: 2019-08-17.

\bibitem[\protect\citeauthoryear{Wu, Mealli, Kioumourtzoglou, Dominici, and
  Braun}{Wu et~al.}{2018}]{wu2018matching}
Wu, X., F.~Mealli, M.~Kioumourtzoglou, F.~Dominici, and D.~Braun (2018).
\newblock Matching on generalized propensity scores with continuous exposures.
\newblock {\em arXiv preprint arXiv:1812.06575\/}.

\bibitem[\protect\citeauthoryear{Zigler, Choirat, and Dominici}{Zigler
  et~al.}{2018}]{zigler2018impact}
Zigler, C.~M., C.~Choirat, and F.~Dominici (2018).
\newblock Impact of national ambient air quality standards nonattainment
  designations on particulate pollution and health.
\newblock {\em Epidemiology (Cambridge, Mass.)\/}~{\em 29\/}(2), 165--174.

\bibitem[\protect\citeauthoryear{Zigler, Dominici, and Wang}{Zigler
  et~al.}{2012}]{zigler2012estimating}
Zigler, C.~M., F.~Dominici, and Y.~Wang (2012).
\newblock Estimating causal effects of air quality regulations using principal
  stratification for spatially correlated multivariate intermediate outcomes.
\newblock {\em Biostatistics\/}~{\em 13\/}(2), 289--302.

\bibitem[\protect\citeauthoryear{Zigler and Papadogeorgou}{Zigler and
  Papadogeorgou}{2018}]{zigler2018bipartite}
Zigler, C.~M. and G.~Papadogeorgou (2018).
\newblock Bipartite causal inference with interference.
\newblock {\em arXiv preprint arXiv:1807.08660\/}.

\end{thebibliography}

\end{document}